\documentclass[preprint,12pt]{elsarticle} 

\usepackage{color}
\usepackage[usenames,dvipsnames]{xcolor}
\usepackage{fullpage}
\usepackage{amsmath}
\usepackage{amssymb}
\usepackage{mathrsfs}
\usepackage{newfloat}
\usepackage{graphicx}
\usepackage{stmaryrd}
\usepackage{subfig}
\usepackage{multirow}
\usepackage{stackengine}
\usepackage{caption}
\usepackage{tikz}
\usepackage{multicol}
\usetikzlibrary{arrows,positioning,calc}
\usepackage{pstool}
\usepackage{textcomp}
\usepackage[hidelinks]{hyperref}
\hypersetup{colorlinks,bookmarksopen,linkcolor=blue,citecolor=blue,urlcolor=blue}
\usepackage{float}
\usepackage{makecell}
\usepackage[export]{adjustbox}
\usepackage{pdfpages}

\definecolor{myred}{RGB}{222,45,38}
\definecolor{myblue}{RGB}{0,115,189}
\definecolor{mygreen}{RGB}{49,156,54}

\newcommand{\bs}[1]{{\boldsymbol{#1}}}
\newcommand{\tr}[1]{\mathrm{tr\,{#1}}}

\def\Aop{\operatornamewithlimits{\mathchoice{\vcenter{\hbox{\huge \sf{A}}}}{\vcenter{\hbox{\Large A}}}{\mathrm{A}}{\mathrm{A}}}}
\DeclareFloatingEnvironment{algorithm}
\DeclareRobustCommand\sampleline[1]{%
	\tikz\draw[#1] (0,0) (0,\the\dimexpr\fontdimen22\textfont2\relax)
	-- (2em,\the\dimexpr\fontdimen22\textfont2\relax);%
}
\newcommand{\mtrx}[1]{{\boldsymbol{\mathsf{#1}}}}
\newcommand{\tensor}[1]{{\boldsymbol{#1}}}

\newcommand{\column}[1]{{\underline{#1}}}
\graphicspath{{figures/}}

\journal{Comput. Methods Appl. Mech. Engrg.}

\biboptions{authoryear}

\begin{document}
	

\begin{frontmatter}
\title{A Newton Solver for Micromorphic Computational Homogenization Enabling Multiscale Buckling Analysis of Pattern-Transforming Metamaterials\tnoteref{titlefoot}}
\tnotetext[titlefoot]{The post-print version of this article is published in \emph{Comput. Methods Appl. Mech. Engrg.}, \href{https://doi.org/10.1016/j.cma.2020.113333}{10.1016/j.cma.2020.113333}}

\author[TUe]{S.E.H.M.~van Bree} 
\ead{s.e.h.m.v.bree@student.tue.nl}

\author[TUe,CTU]{O.~Roko\v{s}\corref{correspondingauthor}} 
\ead{O.Rokos@tue.nl}

\author[TUe]{R.H.J.~Peerlings}
\ead{R.H.J.Peerlings@tue.nl}

\author[CTU]{M.~Do\v{s}k\'{a}\v{r}}
\ead{martin.doskar@fsv.cvut.cz}

\author[TUe]{M.G.D.~Geers}
\ead{M.G.D.Geers@tue.nl}

\address[TUe]{Mechanics of Materials, Department of Mechanical Engineering, Eindhoven University of Technology, P.O.~Box~513, 5600~MB~Eindhoven, The~Netherlands}
\cortext[correspondingauthor]{Corresponding author.}

\address[CTU]{Department of Mechanics, Faculty of Civil Engineering, Czech Technical University in Prague, Th\'{a}kurova~7, 166~29 Prague~6, Czech Republic.}


\begin{abstract}

Mechanical metamaterials feature engineered microstructures designed to exhibit exotic, and often counter-intuitive, effective behaviour such as negative Poisson's ratio or negative compressibility. Such a specific response is often achieved through instability-induced transformations of the underlying periodic microstructure into one or multiple patterning modes. Due to a strong kinematic coupling of individual repeating microstructural cells, non-local behaviour and size effects emerge, which cannot easily be captured by classical homogenization schemes. In addition, the individual patterning modes can mutually interact in space as well as in time, while at the engineering scale the entire structure can buckle globally.
For efficient numerical predictions of macroscale engineering applications, a micromorphic computational homogenization scheme has recently been developed~(\citeauthor{Rokos:2019}, \emph{J. Mech. Phys. Solids}~{\bf 123}, 119--137, 2019). Although this framework is in principle capable of accounting for spatial and temporal interactions between individual patterning modes, its implementation relied on a gradient-based quasi-Newton solution technique. This solver is suboptimal because~(i) it has sub-quadratic convergence, and~(ii) the absence of Hessians does not allow for proper bifurcation analyses. Given that mechanical metamaterials often rely on controlled instabilities, these limitations are serious. Addressing them will reduce the dependency of the solution on the initial guess by perturbing the system towards the correct deformation when a bifurcation point is encountered. Eventually, this enables more accurate and reliable modelling and design of metamaterials. To achieve this goal, a full Newton method, entailing all derivations and definitions of the tangent operators, is provided in detail in this paper.
The construction of the macroscopic tangent operator is not straightforward due to specific model assumptions on the decomposition of the underlying displacement field pertinent to the micromorphic framework, involving orthogonality constraints. Analytical expressions for the first and second variation of the total potential energy are given, and the complete algorithm is listed. 
The developed methodology is demonstrated with two examples in which a competition between local and global buckling exists and where multiple patterning modes emerge. The numerical results indicate that local to global buckling transition can be predicted within a relative error of~$6\%$ in terms of the applied strains. The expected pattern combinations are triggered even for the case of multiple patterns.

\end{abstract}

\begin{keyword}
Mechanical metamaterials \sep computational homogenization \sep micromorphic continuum \sep Newton method \sep bifurcation analysis
\end{keyword}

\end{frontmatter}

%
%
\section{Introduction}
\label{introduction}
Acting like a carefully engineered structure, rather than a standard bulk material, is a common characteristic of mechanical metamaterials. Recent advances in 3D printing and additive manufacturing enable the production of such structures on a relatively small scale, allowing to treat them as a homogeneous medium. Metamaterials are typically designed to exhibit an exotic behaviour which cannot be found in nature, such as a negative compressibility~\citep{Nicolaou:2012}, negative Poisson's ratio~\citep{Kolken:2017}, or a high stiffness with an ultra low density~\citep{Zheng:2014}. In this contribution, we focus on elastomeric mechanical metamaterials, which under compression exhibit microstructural buckling resulting in a pattern transformation. Such a transformation induces an abrupt change in effective properties including Young's modulus and Poisson's ratio, with envisioned applications in, e.g., soft robotics~\cite[see][]{Yang:2015,Mark:2016,Mirzaali:2018}.

\begin{figure}[htbp]
	\centering
	\subfloat[local buckling]{\includegraphics[width=1.8cm,angle=90]{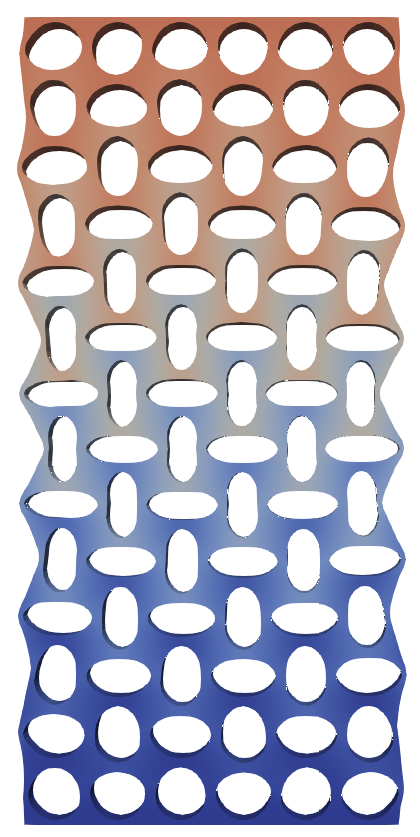}\label{fig:DNSlocalvsglobala}}
	\hspace{1em}
	\subfloat[global buckling]{\includegraphics[width=2.7cm,angle=90]{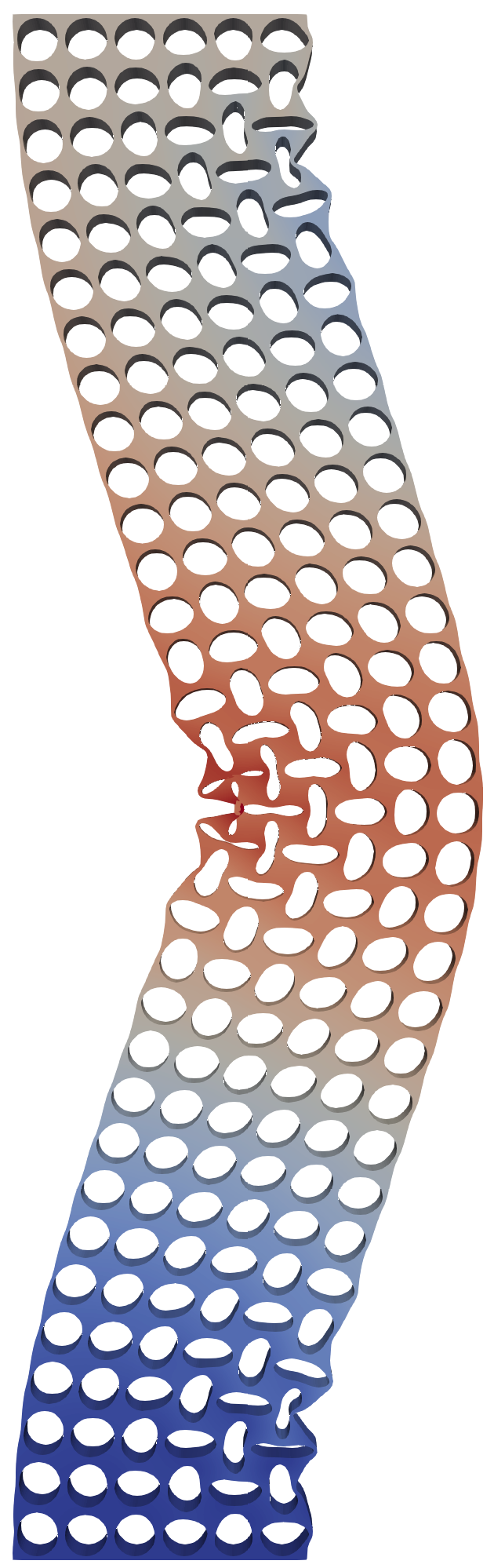}\label{fig:DNSlocalvsglobalb}}
	\hspace{1em}	
	\def\svgwidth{1.25cm}
\begingroup%
  \makeatletter%
  \providecommand\color[2][]{%
    \errmessage{(Inkscape) Color is used for the text in Inkscape, but the package 'color.sty' is not loaded}%
    \renewcommand\color[2][]{}%
  }%
  \providecommand\transparent[1]{%
    \errmessage{(Inkscape) Transparency is used (non-zero) for the text in Inkscape, but the package 'transparent.sty' is not loaded}%
    \renewcommand\transparent[1]{}%
  }%
  \providecommand\rotatebox[2]{#2}%
  \newcommand*\fsize{\dimexpr\f@size pt\relax}%
  \newcommand*\lineheight[1]{\fontsize{\fsize}{#1\fsize}\selectfont}%
  \ifx\svgwidth\undefined%
    \setlength{\unitlength}{375bp}%
    \ifx\svgscale\undefined%
      \relax%
    \else%
      \setlength{\unitlength}{\unitlength * \real{\svgscale}}%
    \fi%
  \else%
    \setlength{\unitlength}{\svgwidth}%
  \fi%
  \global\let\svgwidth\undefined%
  \global\let\svgscale\undefined%
  \makeatother%
  \begin{picture}(1,2)%
    \lineheight{1}%
    \setlength\tabcolsep{0pt}%
    \put(0.68045508,0.89199442){\color[rgb]{0,0,0}\makebox(0,0)[lt]{\lineheight{1.25}\smash{\begin{tabular}[t]{l}\tiny$\|\vec{u}\|_2/u_\mathrm{D}$\end{tabular}}}}%
    \put(0,0){\includegraphics[width=\unitlength,page=1]{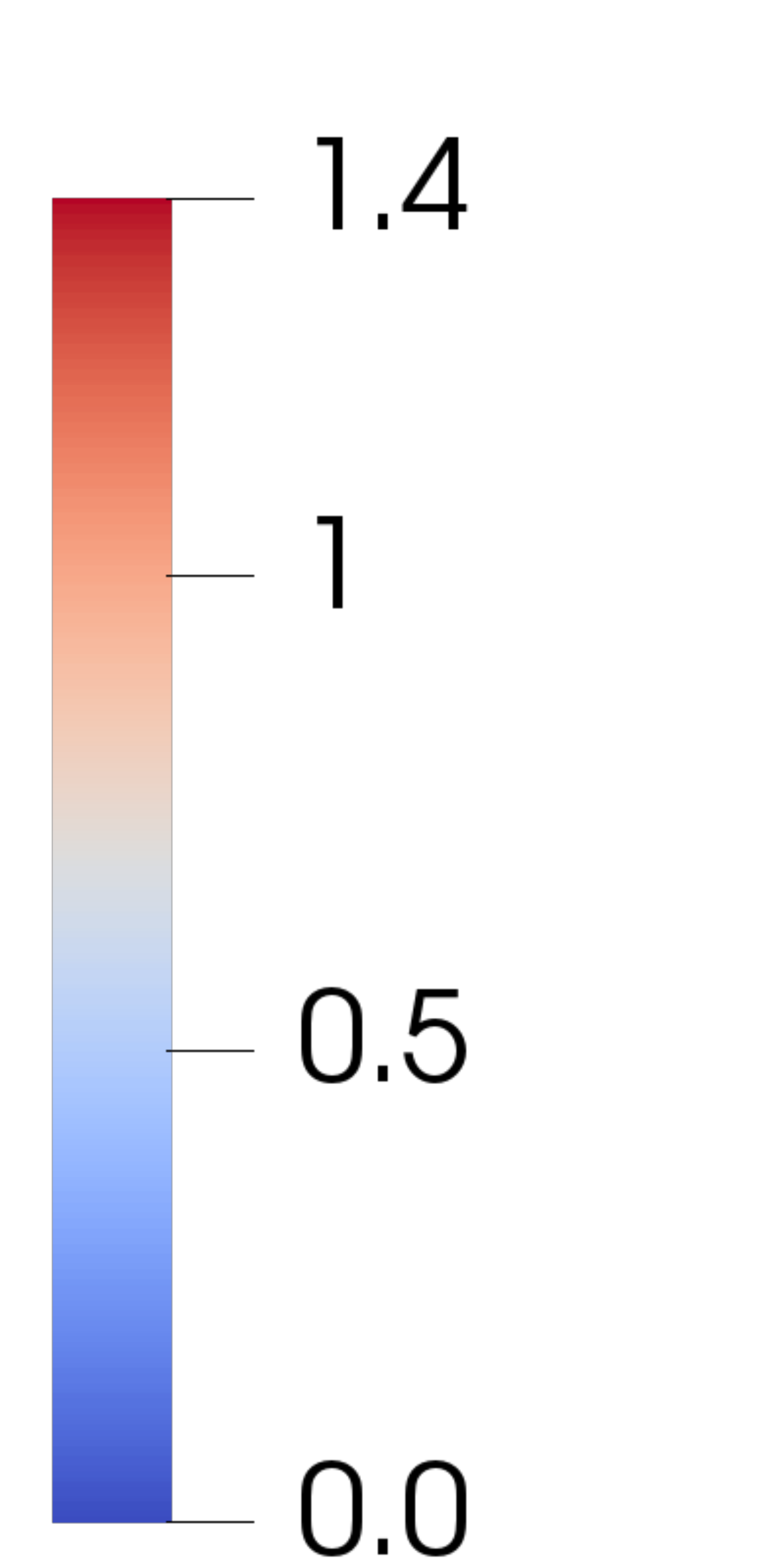}}%
  \end{picture}%
\endgroup%
	
	\caption{An elastomeric mechanical metamaterial containing circular holes in a square packing exhibiting a pattern transformation under a compressive load induced by clamping the right-hand side of the specimen while prescribing a horizontal displacement~$u_\mathrm{D}$ at the left-hand side. Restricted evolution of the patterning near the edges results in stiff boundary layers. Depending on the slenderness of the specimen, local~(a) or global~(b) buckling occurs. In the case of global buckling, the patterning mode (i.e.~local buckling) is localized only in the compressive regions of the specimen. The colour indicates the pointwise magnitude of the displacement field~$\vec{u}$ relative to~$u_\mathrm{D}$.}
	\label{fig:DNSlocalvsglobal}
\end{figure}

Because elastomeric mechanical metamaterials rely mostly on local instabilities in their microstructural morphology, large deformations, rotations, and strains occur. In particular, the microstructure undergoes a pattern transformation, due to coordinated buckling of the underlying microstructure, resulting in a strongly non-local behaviour. If the specimen is restricted, e.g.~by applied essential boundary conditions, the expected pattern cannot fully develop, and in the vicinity of the restriction the so-called boundary layers are formed. Because of pattern restriction such boundary layers generally behave stiffer compared to the bulk of the (meta)material, which may significantly influence the overall response even at the engineering scale. Such a configuration is depicted in Fig.~\ref{fig:DNSlocalvsglobala}, in which an example of an elastomeric metamaterial beam subjected to compressive load is shown. The buckled pattern vanishes close to the two vertical boundaries, resulting in a stiffening effect. The extent to which the boundary layers influence the effective mechanical behaviour depends on the ratio of their thickness and the overall size of the specimen, or more generally on the scale ratio defined as the ratio between the overall size of the specimen~$H$ relative to the typical size of the microstructural features~$\ell$, i.e.~$H/\ell$.

For predictive modelling of engineering-scale applications, it is important to accurately yet efficiently predict the overall mechanical response of the structure. To this end, homogenization techniques are employed, replacing the complex microstructural behaviour with an equivalent continuum model. However, due to the non-locality, patterning, and buckling of the microstructure, homogenization of mechanical metamaterials presents a difficult challenge. First-order computational homogenization, outlined e.g.~by~\cite{Kouznetsova:2001}, can significantly reduce computing time compared to full scale Direct Numerical Simulations~(DNS). However, its inherent assumptions on locality and scale separation prevent it from predicting any size effects in the microstructure. \cite{Ameen:2018} showed that as a consequence relative errors up to~$40\%$ can be induced in terms of force quantities by the first-order method in the post-bifurcation regime for small scale ratios. With an increasing scale ratio, the accuracy of the predicted overall behaviour typically improves, with an exact match for~$H/\ell \rightarrow \infty$. The second-order computational homogenization~\citep{Kouznetsova:2004}, which incorporates the gradient of the macroscopic deformation gradient in the micro-to-macro transition, permits to reflect non-locality and ensuing size-effects. The effective behaviour captured by this method coincides satisfactorily with DNS results even for low scale ratios, albeit at the cost of additional complexity stemming from a higher-order continuum formulation at the macroscale level; see~\cite{Sperling:2019} for more details. Recently, \cite{Rokos:2019} proposed a micromorphic computational homogenization framework, specifically designed to predict the effective behaviour of mechanical metamaterials by decomposing the displacement field into three components: (i)~a smooth, mean displacement field, (ii)~a spatially correlated microfluctuation field, and~(iii) an uncorrelated, local microfluctuation field. This decomposition ensures an adequate performance and accuracy by introducing prior knowledge on the patterning fluctuation. See~\cite{Sperling:2019} and~\cite{Rokos:2019a} for more details.

A major practical limitation of the previously reported implementations of the micromorphic computational homogenization framework is that a quasi-Newton solution method has been employed. This method is not as efficient as a full Newton scheme, it has shown to be quite sensitive to the initialization of the (equilibrium) iteration process, and in particular to the perturbations applied to trigger buckling and, related to the latter point, it does not allow for a proper bifurcation analysis. For elastomeric metamaterials in particular, such a buckling analysis is essential for the prediction of the local patterning (resulting in a transition in the effective mechanical properties, see Fig.~\ref{fig:DNSlocalvsglobala}) or global buckling (indicating a potential failure of the entire structure, see Fig.~\ref{fig:DNSlocalvsglobalb}). Without a proper Newton algorithm, such phenomena can hardly be captured in an accurate and reliable way.

The macroscopic instability at the level of a material point induced by a microscopic bifurcation has been studied by~\cite{Saiki:2002}, whereas~\cite{Wadee:2015} have investigated geometrical effects on the buckling behaviour of cellular structures in which a transition between local and global buckling is observed as a function of the specimen slenderness. Experimentally and numerically, \cite{Niknam:2018} have compared in-plane and out-of-plane buckling of various types and sizes of architected cellular structures. Specimens with a hexagonal honeycomb microstructure, which exhibit multiple buckling patterns under different compressive biaxiality ratios, were studied by \cite{Ohno:2002}. \cite{Rokos:2019a} demonstrated the same multi-pattern character for hexagonally stacked cells with circular holes using the micromorphic homogenization framework. Obtaining correct patterning of the microstructure required, nevertheless, intervention of the user based on insight in the mechanics of the system, precisely because no reliable solver was available to tackle the buckling.

The main goal of this paper is to derive the tangent operator for the micromorphic computational homogenization framework, enabling a more efficient and robust solution procedure using a full Newton algorithm and allowing for bifurcation analyses~\citep{Miehe:2007}. Unlike the derivation of the Hessians (i.e., the macroscopic tangents or stiffnesses) for first-order computational homogenization---see detailed explanation in~\citep{Miehe:2002} or~\citep{Miehe:2003}---, the Hessians for the micromorphic scheme require a non-trivial extension. Additional orthogonality constraints acting within each Representative Volume Element~(RVE) need to be enforced in order to guarantee uniqueness of the adopted kinematic decomposition. Using variational calculus, the first variation of the averaged energy resulting in the microscopic and macroscopic governing equations is derived, followed by the second variation from which the micro-, and coupling macro-Hessians can be obtained. Following~\cite{Rokos:2019a}, the formulation involves an arbitrary number of patterning modes, and introduces a slight reformulation of the orthogonality constraints with respect to gradients of individual modes as compared to the original framework~\citep{Rokos:2019} in order to eliminate spurious oscillations observed in the resulting micromorphic fields. Employing a standard Finite Element~(FE) discretization, the associated internal forces and stiffnesses are constructed, from which the local microfluctuation fields are condensed out, yielding a macroscopic Newton algorithm. We illustrate the performance of the method with two examples, one focusing on local versus global buckling of a metamaterial column, and one on local patterning of a hexagonally-voided microstructure.

The remainder of this paper is organized as follows. After recalling the kinematic decomposition pertinent to the micromorphic framework, Section~\ref{ch:micromorphic} details the derivation of the first and second variation of the ensemble averaged energy, resulting in  both macro- and microscopic governing equations accompanied by the relevant macro- and microscopic Hessians. Employing standard FE procedures, Section~\ref{ch:numerical} describes the discretization of the governing equations at both scales and addresses the bifurcation analysis. Section~\ref{results} illustrates the developed methodology with two examples: (i)~a compressed metamaterial column with a varying slenderness ratio in which a competition between local and global buckling exists, and~(ii) a microstructure with hexagonally-stacked holes subjected to biaxial compressive loading exhibiting three distinct pattern transformations. Finally, the summary and conclusions follow in Section~\ref{ch:conclusions}.

Throughout the paper, the following notational conventions are used
\begin{multicols}{2}
\begin{itemize}[\textbf{-}]
\itemsep0em 
\item scalars~$ a $,
\item vectors~$ \vec{a} $,
\item position vector in the reference configuration~$ \vec{X} = X_1\vec{e}_1 + X_2\vec{e}_2 $,
\item second-order tensors~$ \bs{A}=A_{ij}\vec{e}_i\vec{e}_j $,
\item third-order tensors~$ {{^3}\bs{A}}=A_{ijk}\vec{e}_i\vec{e}_j\vec{e}_k $,
\item fourth-order tensors~$ {{^4}\bs{A}}=A_{ijkl}\vec{e}_i\vec{e}_j\vec{e}_k\vec{e}_l $,
\item matrices~$ \bs{\mathsf{A}} $ and column matrices~$ \underline{a} $,
\item $ \vec{a} \cdot \vec{b} = a_i b_i $,
\item $ \bs{A} \cdot \vec{b} =  A_{ij} b_j\vec{e}_i $,
\item $\bs{A}\cdot\bs{B} = A_{ik}B_{kj}\vec{e}_i\vec{e}_j $,
\item $\bs{A}:\bs{B} = A_{ij}B_{ji}$,
\item transpose~$ \bs{A}^\mathsf{T}$, $ A_{ij}^\mathsf{T} = A_{ji} $,
\item right transpose~${^4}\bs{A}^{\mathsf{RT}},{A}^{\mathsf{RT}}_{ijkl} = {A}_{ijlk} $,
\item left transpose~${^4}\bs{A}^{\mathsf{LT}},{A}^{\mathsf{LT}}_{ijkl} = {A}_{jikl} $,
\item gradient operator~$  \vec{\nabla} \vec{a} = \textstyle \frac{\partial a_j}{\partial X_i} \vec{e}_i \vec{e}_j $,
\item divergence operator~$  \vec{\nabla} \cdot \vec{a} = \textstyle \frac{\partial a_i}{\partial X_i} $,
\item integration~$\langle{f}(\vec{X}_\mathrm{m})\rangle_{\Omega_\bullet} = \int_{\Omega_\bullet} f(\vec{X}_\mathrm{m})\, \mathrm{d}\vec{X}_\mathrm{m}$,
\item derivatives of scalar functions with respect to second-order tensors \\
$\displaystyle \delta\Psi(\bs{F};\delta\bs{F}) = \left.\frac{\mathrm{d}}{\mathrm{d}h}\Psi(\bs{F}+h\delta\bs{F})\right|_{h=0} = \frac{\partial\Psi(\bs{F})}{\partial\bs{F}}:\delta\bs{F} $,
\end{itemize}
\end{multicols}
\noindent
where Einstein's summation convention is adopted on repeated indices~$i$, $j$, $k$, $l$, and~$\vec{e}_i$, $i = 1, 2,$ denote the basis vectors of a two-dimensional Cartesian coordinate frame.
%
%
\section{Reformulation of the Micromorphic Computational Homogenization Framework and Derivation of the Tangents}
\label{ch:micromorphic}
%
%
\subsection{Kinematic Decomposition}
\begin{figure}
	\centering
	\begin{tikzpicture}[x = 0.77cm, y = 0.77cm]
	\coordinate (A) at (-2,1.5);
	\coordinate (B) at (4,1.5);
	\coordinate (C) at (4,4.5);
	\coordinate (D) at (-2,4.5);
	\filldraw[draw = white, fill = blue!5] (A) -- (B) -- (C) -- (D) -- cycle;
	\draw[thin] (D) -- (C) node[anchor = north east] {\footnotesize $\Omega$};
	\draw[thin] (A) -- (B);
	\draw[thin] (A) -- (D);
	\draw[thin] (B) -- (C);
	\draw[-latex',line width=0.3mm] (-2,1.5) -- (-2,3) node[anchor =  east] {\footnotesize $\vec{e}_{2}$};
	\draw[-latex',line width=0.3mm] (-2,1.5) -- (-0.5,1.5) node[anchor = north] {\footnotesize $\vec{e}_{1}$};
	\draw[-latex',dashed] (-2,1.5) -- (1,3) node[anchor = north,shift={(-0.7,-0.3)}]{\footnotesize $\vec{X}$};
	\filldraw (1,3) circle(0.05);
	\draw (1.15,5) node {\footnotesize Macroscale boundary value problem, Eq.~\eqref{eq:Mbalanceequations}};
	\draw[-latex',gray] (1.4,3) .. controls (2.4,2.6) and (3.5,1.3) .. (4,-0.5);
	\draw (4.2,0.8) node[anchor = west] {\footnotesize $\vec{\nabla}\vec{v_0}$};
	\draw (4.4,0.1) node[anchor = west] {\footnotesize $v_i, \, \vec{\nabla}{v_i}$};
	\filldraw[fill  = blue!5] (2.5,-0.6) -- (5.5,-0.6) -- (5.5, -3.6) -- (2.5, -3.6) -- cycle;
	\foreach \x in {3.25,4.75}
	\foreach \y in {-1.35,-2.85}
	\filldraw[draw = black, fill = white] (\x,\y) circle(0.65);
	\draw[-latex',line width=0.3mm] (4,-2.1) -- (4,-1.4) node[anchor =  east,shift={(-0.1,0)}] {\footnotesize $\vec{e}_{2\mathrm{m}}$};
	\draw[-latex',line width=0.3mm] (4,-2.1) -- (4.8,-2.1) node[anchor = north,shift={(0,-0.15)}] {\footnotesize $\vec{e}_{1\mathrm{m}}$};
	\draw[-latex',dashed] (4,-2.1) -- (5.0,-0.67) node[anchor = north west,shift={(-0.6,-0.3)}]{\footnotesize $\vec{X}_\mathrm{m}$};
	\node[anchor = north east] at (4.65,-3) {\footnotesize $\Omega_\mathrm{m}$};
	\filldraw (5.0,-0.67) circle(0.05);
	\draw[-latex',gray](-2,-0.5) .. controls (-1.5,1.3) and (-0.4,2.6) .. (0.6,3);
	\draw (-4.5,0.8) node[anchor=west] {\footnotesize $\boldsymbol{\Theta}, \, \Pi_i, \, \vec{\Lambda}_i$};
	\draw (-4.8,0.1) node[anchor=west] {\footnotesize Stiffnesses};
	\filldraw[fill = blue!5] (-0.5,-0.6) .. controls (-1.5,-1.1) and (-2.5,-0.1) .. (-3.5,-0.6) .. controls (-3,-1.6) and (-4,-2.6) .. (-3.5, -3.6) .. controls (-2.5,-3.1) and (-1.5,-4.1) .. (-0.5, -3.6) .. controls (-1,-2.6) and (0,-1.6) .. (-0.5,-0.6);
	\filldraw[draw = black, fill = white] (-2.75,-1.35) ellipse(0.5 and 0.75);
	\filldraw[draw = black, fill = white] (-1.25,-1.35) circle(0.75 and 0.5);
	\filldraw[draw = black, fill = white] (-2.75,-2.85) circle(0.75 and 0.5);
	\filldraw[draw = black, fill = white] (-1.25,-2.85) circle(0.5 and 0.75);
	\draw[-latex'](2.4,-2.3) .. controls (1.6,-1.5) and (0.4,-1.5) .. (-0.4,-2.3);
	\draw (1,-2.5) node {\footnotesize Solve for~$\vec{w}$};	
	\draw (1,-4.25) node {\footnotesize Microscale boundary value problem, Eq.~\eqref{eq:mgoverning}};
	\end{tikzpicture}
	\caption{A schematic representation of the micromorphic computational homogenization framework. The macroscopic displacement gradient~$\vec{\nabla}\vec{v}_0$, micromorphic fields~$v_i$, and their spatial gradients~$\vec{\nabla}v_i$ at a macroscopic point~$\vec{X}$ are sampled and sent to the microscale, where a boundary value problem for the microfluctuation field~$\vec{w}$ is solved. Based on the solution~$\vec{w}$, the macroscopic stresses, $\boldsymbol{\Theta}$, $\Pi_i$, $\vec{\Lambda}_i$, and stiffnesses are computed and passed back to the macroscale, where the macroscopic boundary value problem is assembled and solved.}
	\label{fig:comphomog}
\end{figure}
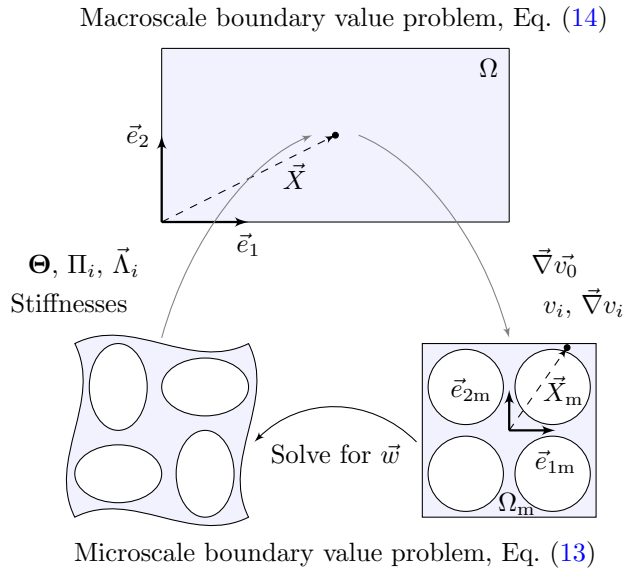

The micromorphic computational homogenization framework~\citep{Rokos:2019}, depicted schematically in Fig.~\ref{fig:comphomog}, relies on the decomposition of the kinematic field~$\vec{u}$ into the mean effective displacement~$ \vec{v}_0 $, long range correlated fluctuation components~$ v_i \vec{\varphi}_i $, $ i = 1, \dots, n $, and the remaining local microfluctuation field~$ \vec{w} $, i.e.
\begin{equation}
\vec{{u}}(\vec{X}) = \vec{v}_0(\vec{X}) + 
\sum_{i=1}^{n} v_i(\vec{X}) \vec{\varphi}_i(\vec{X}) + 
\vec{w}(\vec{X}).
\label{eq:u}
\end{equation}
The vector field~$ \vec{\varphi}_i $ corresponds to the~$ i $-th patterning mode of the underlying microstructure, whereas the scalar field~$ v_i $ regulates, spatially and in time, its magnitude. Unlike the original formulation~\citep{Rokos:2019}, which considered only one such mode, we consider here an arbitrary number of modes, $n$; see also~\cite{Rokos:2019a}. Because in general it may not be possible to control the positioning of the microstructure relative to the specimen's boundary, all possible microstructural translations should be taken into account via ensemble averaging~\citep[cf.][]{Ameen:2018}. The micromorphic scheme avoids this costly procedure by approximating the mechanical state of a point in a translated microstructure by evaluating the mechanical state of a microstructurally equivalent point in the reference microstructure. In addition, a separation of scales into a macroscopic position vector~$ \vec{X} $ and a microscopic position vector~$ \vec{X}_\mathrm{m} $ is introduced, assuming the fields~$\vec{v}_0$ and~$v_i$ to vary slowly over a close vicinity of each macroscopic point~$\vec{X}$ spanned by a microscopic Representative Volume Element~(RVE) with a domain~$\Omega_\mathrm{m}$. Consequently, the microfluctuation field~$\vec{w}$ is computed only locally over each RVE, and is independent for RVEs associated with distinct macroscopic points~$\vec{X}$; they communicate only by means of the macroscopic fields~$\vec{v}_0$ and~$v_i$. Therefore, $ \vec{v}_0 $ and~$ v_i $ become functions of the macroscopic position only, whereas~$ \vec{\varphi}_i $ and~$ \vec{w} $ are functions of the microscopic as well as the macroscopic position. Because the patterning mode~$\vec{\varphi}_i$ is the same for each macroscopic point, it eventually depends on the microscopic position vector~$\vec{X}_\mathrm{m}$ only. Using a first-order Taylor expansion and the above considerations, the decomposition of Eq.~\eqref{eq:u} can be approximated as
\begin{equation}
\begin{aligned}
 \vec{{u}}(\vec{X},\vec{X}_\mathrm{m}) &\approx \vec{v}_0(\vec{X}) + \vec{X}_\mathrm{m}\cdot\vec{\nabla}\vec{v}_0(\vec{X}) \\
&+
\sum_{i=1}^{n} [v_i(\vec{X}) + \vec{X}_\mathrm{m}\cdot\vec{\nabla}v_i(\vec{X})]\vec{\varphi}_i(\vec{X}_\mathrm{m}) \\
&+
\vec{w}(\vec{X},\vec{X}_\mathrm{m}), \quad \vec{X} \in \Omega, \vec{X}_\mathrm{m} \in \Omega_\mathrm{m}. 
\end{aligned}
\label{eq:approxu}
\end{equation}
For more details on the decomposition~\eqref{eq:approxu}, the reader is referred to~\cite{Rokos:2019}---albeit for a single mode~$\vec{\varphi}_1$. The fields~$ \vec{v}_0 $, $ v_i $, and~$ \vec{w} $, are unknown and need to be solved for, while the individual patterning modes~$ \vec{\varphi}_i $ are characteristic for the underlying microstructural morphology and are assumed to be known a priori, computed either from a Bloch-type analysis~\citep{Bertoldi:2008}, estimated analytically from full-scale numerical simulations~\citep{Rokos:2019}, or identified experimentally~\citep{Siavash:2020}. Because the patterning modes~$\vec{\varphi}_i$ are defined with respect to the reference microscopic configuration~$\vec{X}_\mathrm{m}$, the effect of the macroscopic rotations needs to be factored out from the macroscopic deformation gradients, $\tensor{F}_\mathrm{M} = \bs{I}+(\vec{\nabla}\vec{v}_0)^\mathsf{T}$, using the polar decomposition in Eq.~\eqref{eq:approxu}, i.e.~$\tensor{F}_\mathrm{M} = \tensor{R}_\mathrm{M}\cdot\tensor{U}_\mathrm{M}$, where~$\tensor{R}_\mathrm{M}$ is the macroscopic rotation tensor and~$\tensor{U}_\mathrm{M}$ is the macroscopic stretch tensor. This effectively means that the term~$\vec{\nabla}\vec{v}_0$ needs to be replaced with~$\tensor{U}_\mathrm{M} - \tensor{I}$ in Eq.~\eqref{eq:approxu}, and that all effective stress and stiffness quantities need to be rotated back accordingly, see e.g.~\cite[][Section~3.2]{Kunc2019} for more details. Such a distinction is, however, omitted hereafter to simplify all derivations, and can even be neglected in the limit of small rotations as is the case for both examples shown in Section~\ref{results}. Note also that the additional micromorphic fields~$v_i$ in Eqs.~\eqref{eq:u} and~\eqref{eq:approxu} relate directly to the displacement field~$\vec{u}$. Such a setup contrasts with standard micromorphic formulations, in which the microdeformation~$\tensor{\chi}$ typically relates to the gradient of~$\vec{u}$, see e.g.~\cite[][Eqs.~(29)--(32)]{Forest2011}. The decomposition of Eq.~\eqref{eq:approxu} further suggests that when all micromorphic fields~$v_i$ vanish, the ansatz of the standard first-order computational homogenization is recovered, i.e.~$\vec{u}(\vec{X},\vec{X}_\mathrm{m}) = \vec{v}_0(\vec{X}) + \vec{X}_\mathrm{m}\cdot\vec{\nabla}\vec{v}_0(\vec{X}) + \vec{w}(\vec{X},\vec{X}_\mathrm{m})$, cf. e.g.~\cite[][Eq.~(1)]{Geers2010}.

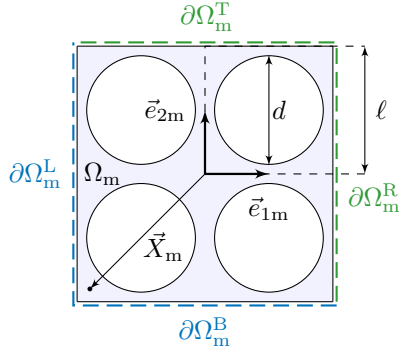
\begin{figure}
	\centering
	\begin{tikzpicture}
		\def\W{2} 
		\def\H{2} 
		\def\cellsize{1.7} 
		\def\diameter{0.85*\cellsize} 
		\coordinate (A) at (0,0);
		\coordinate (B) at (\cellsize*\W,0);
		\coordinate (C) at (\cellsize*\W,\cellsize*\H);
		\coordinate (D) at (0,\cellsize*\H);
		\filldraw[draw = white, fill = blue!5] (A) -- (B) -- (C) -- (D) -- cycle;
		\draw[thin] (D) -- (C);
		\draw[thin] (A) -- (B);
		\draw[thin] (A) -- (D);
		\draw[thin] (B) -- (C);
		\def\skippbc{0.05}
		\draw[thick,mygreen,dash pattern=on 6pt off 3pt] (\cellsize*\W+\skippbc,0) -- (\cellsize*\W+\skippbc,\cellsize*\H+\skippbc) node[midway,below right] {\footnotesize $\partial\Omega_\mathrm{m}^\mathrm{R}$} -- (0,\cellsize*\H+\skippbc) node[midway,above] {\footnotesize $\partial\Omega_\mathrm{m}^\mathrm{T}$};
		\draw[thick,myblue,dash pattern=on 6pt off 3pt] (\cellsize*\W+\skippbc,-\skippbc) -- (-\skippbc,-\skippbc) node[midway,below] {\footnotesize $\partial\Omega_\mathrm{m}^\mathrm{B}$} -- node[midway,left] {\footnotesize $\partial\Omega_\mathrm{m}^\mathrm{L}$} (-\skippbc,\cellsize*\H+\skippbc);	
		\draw (0.2*\cellsize,\cellsize) node{\footnotesize $\Omega_\mathrm{m}$};
		\foreach \i in {1,...,\W}{%
			\foreach \j in {1,...,\H}{%
				\filldraw[fill=white,thin] ({\cellsize*(\i-0.5)},{\cellsize*(\j-0.5)}) circle ({\diameter/2});
		}}
		\draw[thin,latex'-latex'] ({\cellsize*(\W+0.25)},\cellsize) -- ({\cellsize*(\W+0.25)},2*\cellsize) node[midway,right] {\footnotesize $\ell$};
		\draw[dashed] ({\cellsize*(\W+0.25)},\cellsize) -- ({\cellsize*(\W-1)},\cellsize) -- ({\cellsize*(\W-1)},2*\cellsize) -- ({\cellsize*(\W+0.25)},2*\cellsize);
		\draw[thin,latex'-latex'] ({\cellsize*(\W-0.5)},{\cellsize*(1+0.0665)}) -- ({\cellsize*(\W-0.5)},{\cellsize*(2-0.0665)}) node[midway,right,shift={(-0.1,0)}] {\footnotesize $d$};
		\draw[-latex',line width=0.3mm] (\cellsize,\cellsize) -- (\cellsize,1.5*\cellsize) node[anchor = east,shift={(-0.1,-0.0)}] {\footnotesize $\vec{e}_{2\mathrm{m}}$};
		\draw[-latex',line width=0.3mm] (\cellsize,\cellsize) -- (1.5*\cellsize,\cellsize) node[anchor = north,shift={(-0.0,-0.15)}] {\footnotesize $\vec{e}_{1\mathrm{m}}$};
		\draw[-latex'] (\cellsize,\cellsize) -- (0.1*\cellsize,0.1*\cellsize) node[anchor = west,midway,shift={(-0.2,-0.2)}]{\footnotesize $\vec{X}_\mathrm{m}$};
		\filldraw (0.1*\cellsize,0.1*\cellsize) circle(0.025);
	\end{tikzpicture}	
	\caption{A schematic example of a pattern-transforming~$2 \times 2$ RVE with a square stacking of holes. The four boundary edges are split into two mirror parts~$\partial\Omega_\mathrm{m}^+ = \partial\Omega_\mathrm{m}^\mathrm{T} \cup \partial\Omega_\mathrm{m}^\mathrm{R}$ and~$\partial\Omega_\mathrm{m}^- = \partial\Omega_\mathrm{m}^\mathrm{B} \cup \partial\Omega_\mathrm{m}^\mathrm{L}$ for the implementation of periodic boundary conditions.}
	\label{fig:RVE}
\end{figure}

The uniqueness of the decomposition~\eqref{eq:approxu} within an RVE is guaranteed by introducing the following additional orthogonality conditions:
\begin{align}
\big\langle \vec{w}(\vec{X},\vec{X}_\mathrm{m}) \big\rangle_{\Omega_\mathrm{m}} &= \vec{0}, \label{eq:constraintsa}\\
\big\langle \vec{w}(\vec{X},\vec{X}_\mathrm{m}) \cdot \vec{\varphi}_i(\vec{X}_\mathrm{m}) \big\rangle_{\Omega_\mathrm{m}} &= 0, \quad i=1,\dots,n,\label{eq:constraintsb}\\
\big\langle \vec{w}(\vec{X},\vec{X}_\mathrm{m}) \cdot [ \vec{\varphi}_i(\vec{X}_\mathrm{m}) \vec{X}_\mathrm{m} ] \big\rangle_{\Omega_\mathrm{m}} &= \vec{0}, \quad i=1,\dots,n.\label{eq:constraintsc}
\end{align}
Recall that throughout this contribution the angle brackets indicate the integration over the domain specified by the subscript. The first condition~\eqref{eq:constraintsa} requires zero mean of~$\vec{w}$ over~$\Omega_\mathrm{m}$, effectively eliminating rigid body translations. The second condition establishes the uniqueness in terms of the patterning modes~$\vec{\varphi}_i$ themselves, because, upon assuming a homogeneous state with~$\vec{\nabla}v_i = \vec{0}$, for instance, the patterning can be equally well represented by the product of the micromorphic field and the patterning mode, i.e.~$v_i\vec{\varphi}_i$, or by the microfluctuation field~$\vec{w}$ while keeping~$v_i = 0$. The last orthogonality condition has not been previously introduced in the original formulation of~\cite{Rokos:2019}. It follows, however, from the decomposition outlined in Eg.~\eqref{eq:approxu}, and eliminates non-uniqueness issues related to linearly varying magnitudes of the patterning fields~$\vec{\varphi}_i$, which may cause spurious macroscopic oscillations for microstructures with a hexagonal stacking of holes. This condition acts mainly as a stabilization condition and does not significantly affect the overall mechanical response. Following the same reasoning as in first-order computational homogenization, the remaining orthogonality condition, which might arise from the non-uniqueness related to the ansatz~\eqref{eq:approxu} (i.e.~orthogonality with respect to~$\vec{X}_\mathrm{m}\cdot\vec{\nabla}\vec{v}_0(\vec{X})$, or~$\big\langle\vec{w}(\vec{X},\vec{X}_\mathrm{m})\vec{X}_\mathrm{m}\big\rangle_{\Omega_\mathrm{m}} = \bs{0}$ for an arbitrary~$\vec{\nabla}\vec{v}_0$), is accounted for differently. As a well-accepted modelling choice, which provides accurate results in the first- as well as second-order computational homogenization schemes, a periodicity constraint on~$\vec{w}$ is adopted ensuring this orthogonality, i.e.
\begin{equation}
\llbracket \vec{w}(\vec{X},\vec{X}_\mathrm{m}) \rrbracket = \vec{0}, \quad \vec{X}_\mathrm{m} \in \partial\Omega_\mathrm{m}^+,
\label{eq:periodicity}
\end{equation}
where~$\llbracket \vec{w}(\vec{X},\vec{X}_\mathrm{m}) \rrbracket = \vec{w}(\vec{X},\partial\Omega_\mathrm{m}^+) - \vec{w}(\vec{X},\partial\Omega_\mathrm{m}^-)$ denotes the jump of the field~$\vec{w}(\vec{X},\vec{X}_\mathrm{m})$ on the RVE boundary split into two parts: $\partial\Omega_\mathrm{m} = \partial\Omega_\mathrm{m}^+ \cup \partial\Omega_\mathrm{m}^-$. As an example, a~$2 \times 2$ RVE of a square stacking of holes is schematically shown in Fig.~\ref{fig:RVE}, where~$\partial\Omega_\mathrm{m}^+ = \partial\Omega_\mathrm{m}^\mathrm{T} \cup \partial\Omega_\mathrm{m}^\mathrm{R}$ and~$\partial\Omega_\mathrm{m}^- = \partial\Omega_\mathrm{m}^\mathrm{B} \cup \partial\Omega_\mathrm{m}^\mathrm{L}$. The same approach is used for polygons with multiple edges---see e.g.~ahead to Fig.~\ref{fig:hexab} for the case of a hexagonal RVE. The periodicity constraint in combination with the condition of Eq.~\eqref{eq:constraintsa} also eliminates all rigid body modes of~$\vec{w}$. Although periodic boundary conditions for~$\vec{w}$ (and thus antiperiodic RVE boundary tractions) are adopted, such a choice might not necessarily result in the optimal performance of the proposed micromorphic scheme. See e.g.~\cite{Forest2011}, where antiperiodic microfluctuation fields~$\vec{w}$ with periodic RVE boundary tractions have been observed. Yet, it will be demonstrated in the results Section~\ref{results} that the periodicity constraints~\eqref{eq:periodicity} provide an adequate accuracy here.
%
%
\subsection{Potential Energy}
Because we are restricting ourselves to hyperelastic materials, the unknown parts~$ \vec{v}_0(\vec{X}) $, $ v_i(\vec{X}) $, and $ \vec{w}(\vec{X},\vec{X}_\mathrm{m}) $ of the solution~$\vec{u}(\vec{X},\vec{X}_\mathrm{m})$ according to~\eqref{eq:approxu} can be found by minimizing the total potential energy of the system while accounting for the constraints introduced above, i.e.
\begin{equation}
( \vec{v}_0(\vec{X}),v_1(\vec{X}),\dots,v_n(\vec{X}),\vec{w}(\vec{X},\vec{X}_\mathrm{m}) )
\in
\mathrm{arg}\ \underset{\vec{u}(\vec{X},\vec{X}_\mathrm{m})}{\mbox{min}} \ \underset{
\scriptsize
\begin{array}{c}
\vec{\mu}(\vec{X}),\ \underline{\nu}(\vec{X}) \\
\underline{\eta}(\vec{X}),\ \vec{\lambda}(\vec{X},\vec{X}_\mathrm{m}) \\
\end{array}
}{\mbox{max}} \ \mathcal{L}(\vec{u},\vec{\mu},\underline{\nu},\underline{\eta},\vec{\lambda}),
\label{micromorphic:eq7}	
\end{equation}
where the displacement field~$\vec{u}(\vec{X},\vec{X}_\mathrm{m})$ is, through Eq.~\eqref{eq:approxu}, also a function of all unknown macroscopic, $\vec{v}_0(\vec{X})$, $v_1(\vec{X})$, $\dots$, $v_n(\vec{X})$, and microscopic, $\vec{w}(\vec{X},\vec{X}_\mathrm{m})$, quantities in addition to the two spatial variables~$\vec{X}$ and~$\vec{X}_\mathrm{m}$, and where the fields~$\vec{\mu}(\vec{X})$, $\underline{\nu}(\vec{X}) = [\nu_1(\vec{X}), \dots, \nu_n(\vec{X})]$, and~$\underline{\eta}(\vec{X}) = [\vec{\eta}_1(\vec{X}), \dots, \vec{\eta}_n(\vec{X})]$ collect the Lagrange multipliers pertinent to rigid body modes and orthogonality constraints with respect to individual patterning modes as their components. While the bulk constraints of Eqs.~\eqref{eq:constraintsa}--\eqref{eq:constraintsc} are considered for each RVE separately, i.e.~$\vec{\mu}(\vec{X})$, $\underline{\nu}(\vec{X})$, and~$\underline{\eta}(\vec{X})$ are a function of the macroscopic position vector~$\vec{X}$ only, the periodicity constraint~\eqref{eq:periodicity} is a continuous function on the boundary of each RVE and therefore, it is also a function of the microscopic position vector~$\vec{X}_\mathrm{m}$, i.e.~$\vec{\lambda}(\vec{X},\vec{X}_\mathrm{m})$. In Eq.~\eqref{micromorphic:eq7}, the Lagrangian~$\mathcal{L} = \mathcal{E} + \mathcal{C}$ consists of the total potential energy~$\mathcal{E}$
\begin{equation}
\mathcal{E}(\vec{u}(\vec{X},\vec{X}_\mathrm{m})) = \frac{1}{|\Omega_\mathrm{m}|}\Big\langle \big\langle \Psi(\vec{X}_\mathrm{m}, \bs{F}) \big\rangle_{\Omega_\mathrm{m}} \Big\rangle_\Omega
\label{eq:energy_definition}
\end{equation}
and the constraint term~$\mathcal{C}$
\begin{equation}
\begin{aligned}
\mathcal{C}(\vec{u}(\vec{X}&,\vec{X}_\mathrm{m}),\vec{\mu}(\vec{X}),\underline{\nu}(\vec{X}),\underline{\eta}(\vec{X}),\vec{\lambda}(\vec{X},\vec{X}_\mathrm{m})) = \frac{1}{|\Omega_\mathrm{m}|}\Big\langle
\vec{\mu}(\vec{X}) \cdot \big\langle \vec{w}(\vec{X},\vec{X}_\mathrm{m}) \big\rangle_{\Omega_\mathrm{m}} \\
&+ \sum_{i=1}^{n} \nu_i(\vec{X}) \big\langle \vec{w}(\vec{X},\vec{X}_\mathrm{m}) \cdot \vec{\varphi}_i(\vec{X}_\mathrm{m})\big\rangle_{\Omega_\mathrm{m}} 
+ \sum_{i=1}^{n} \vec{\eta}_i(\vec{X}) \cdot \big\langle \vec{w}(\vec{X},\vec{X}_\mathrm{m}) \cdot [ \vec{\varphi}_i(\vec{X}_\mathrm{m}) \vec{X}_\mathrm{m} ]\big\rangle_{\Omega_\mathrm{m}} \\
&- \big\langle\vec{\lambda}(\vec{X},\vec{X}_\mathrm{m})\cdot \llbracket \vec{w}(\vec{X},\vec{X}_\mathrm{m}) \rrbracket \big\rangle_{\partial\Omega_\mathrm{m}^+}
\Big\rangle_\Omega,
\end{aligned}	
\label{eq:constraint_definition}
\end{equation}
where the minus sign in front of the last constraint term is adopted for consistency reasons~\cite[][Section~2.2.2]{Miehe:2007}, and the remaining terms take a plus sign to obtain positive quantities on the right-hand side of the resulting governing equation (see Eq.~\eqref{eq:mgoverning} below). Note that this slightly deviates from~\cite{Rokos:2019}, where the constraint term~$\mathcal{C}$ was incomplete. A hyperelastic constitutive law specified through the energy density function~$\Psi(\vec{X}_\mathrm{m},\bs{F})$ is adopted for the description of the material behaviour; $\bs{F}(\vec{{u}}(\vec{X},\vec{X}_\mathrm{m})) = \bs{I} + (\vec{\nabla}_\mathrm{m}\vec{u}(\vec{X}, \vec{X}_\mathrm{m}))^\mathsf{T}$ is the deformation gradient and~$\vec{\nabla}_\mathrm{m} = \vec{e}_i\,\partial/\partial X_{\mathrm{m},i}$ is the microscopic gradient operator defined in the reference configuration. Note that the explicit dependency of~$\bs{F}$ on~$\vec{u}$ has been dropped in Eq.~\eqref{eq:energy_definition}, and will be omitted for brevity hereafter as well.
%
%
\subsection{First Variation and Governing Equations}
A minimizer of the total potential energy can be found by taking the G\^{a}teaux derivative (i.e.~the first variation) of the Lagrangian~$\mathcal{L}$ and requiring it to vanish:
\begingroup
\allowdisplaybreaks
\begin{align}
0 &= \delta \mathcal{L}(\vec{u},\vec{\mu},\underline{\nu},\underline{\eta},\vec{\lambda};\delta\vec{u},\delta\vec{\mu},\delta\underline{\nu},\delta\underline{\eta},\delta\vec{\lambda}) = \frac{1}{|\Omega_\mathrm{m}|}\Big\langle
\big\langle\bs{P}(\vec{X}_\mathrm{m},\bs{F}):\vec{\nabla}_\mathrm{m}\delta\vec{u}(\vec{X},\vec{X}_\mathrm{m})\big\rangle_{\Omega_\mathrm{m}} \nonumber \\
&+ \vec{\mu}(\vec{X}) \cdot \big\langle \delta\vec{w}(\vec{X},\vec{X}_\mathrm{m}) \big\rangle_{\Omega_\mathrm{m}} + \delta\vec{\mu}(\vec{X}) \cdot \big\langle \vec{w}(\vec{X},\vec{X}_\mathrm{m}) \big\rangle_{\Omega_\mathrm{m}} \nonumber \\
&+ \sum_{i=1}^{n} \nu_i(\vec{X}) \, \big\langle \delta\vec{w}(\vec{X},\vec{X}_\mathrm{m}) \cdot \vec{\varphi}_i(\vec{X}_\mathrm{m}) \big\rangle_{\Omega_\mathrm{m}} +  \sum_{i=1}^{n} \delta\nu_i(\vec{X}) \, \big\langle \vec{w}(\vec{X},\vec{X}_\mathrm{m}) \cdot \vec{\varphi}_i(\vec{X}_\mathrm{m}) \big\rangle_{\Omega_\mathrm{m}} \label{eq:firstvariation} \\
&+ \sum_{i=1}^{n} \vec{\eta}_i(\vec{X}) \cdot \big\langle \delta\vec{w}(\vec{X},\vec{X}_\mathrm{m}) \cdot [ \vec{\varphi}_i(\vec{X}_\mathrm{m})\vec{X}_\mathrm{m} ] \big\rangle_{\Omega_\mathrm{m}}
+\sum_{i=1}^{n} \delta\vec{\eta}_i(\vec{X})\cdot \big\langle \vec{w}(\vec{X},\vec{X}_\mathrm{m}) \cdot [ \vec{\varphi}_i(\vec{X}_\mathrm{m})\vec{X}_\mathrm{m} ] \big\rangle_{\Omega_\mathrm{m}} \nonumber \\
&- \big\langle \vec{\lambda}(\vec{X},\vec{X}_\mathrm{m})\cdot \llbracket \delta\vec{w}(\vec{X},\vec{X}_\mathrm{m}) \rrbracket + \delta\vec{\lambda}(\vec{X},\vec{X}_\mathrm{m})\cdot \llbracket \vec{w}(\vec{X},\vec{X}_\mathrm{m}) \rrbracket \big\rangle_{\partial\Omega_\mathrm{m}^+}
\Big\rangle_{\Omega}, \nonumber
\end{align}
\endgroup
where the local first Piola--Kirchhoff stress tensor~${\bs{P}(\vec{X}_\mathrm{m},\bs{F})}$ is defined as
\begin{equation}
\bs{P}(\vec{X}_\mathrm{m},\bs{F}(\vec{u}(\vec{X},\vec{X}_\mathrm{m}))) = \frac{\partial\Psi(\vec{X}_\mathrm{m},\bs{F}(\vec{u}(\vec{X},\vec{X}_\mathrm{m})))}{\partial\bs{F}^\mathsf{T}}.
\label{eq:stress}
\end{equation}
Making use of the decomposition introduced in Eq.~\eqref{eq:approxu}, $\vec{\nabla}_\mathrm{m}\delta\vec{u}$ reads
\begin{equation}
\begin{aligned}
\vec{\nabla}_\mathrm{m}\delta\vec{{ {u}}}(\vec{X},\vec{X}_\mathrm{m}) &= \vec{\nabla}\delta\vec{v}_0(\vec{X})+ \sum_{i=1}^{n} \vec{\nabla}\delta v_i(\vec{X})\vec{\varphi}_i(\vec{X}_\mathrm{m})\\
&+
\sum_{i=1}^{n} \big[\delta v_i(\vec{X})+\vec{X}_\mathrm{m}\cdot\vec{\nabla}\delta v_i(\vec{X})\big]\vec{\nabla}_\mathrm{m}\vec{\varphi}_i(\vec{X}_\mathrm{m})+\vec{\nabla}_\mathrm{m}\delta\vec{w}(\vec{X},\vec{X}_\mathrm{m}).
\end{aligned}
\label{eq:du}
\end{equation}
Upon substituting this expansion, application of the divergence theorem, and rearrangement of individual terms, Eq.~\eqref{eq:firstvariation} leads to a set of microscopic and macroscopic balance equations. At the microscale, only the variations of the microfluctuation field $\vec{w}(\vec{X},\vec{X}_\mathrm{m})$ and its related Lagrange multipliers~$\vec{\mu}(\vec{X})$, $\nu_i(\vec{X})$, $\vec{\eta}_i(\vec{X})$, and~$\vec{\lambda}(\vec{X},\vec{X}_\mathrm{m})$ matter. Since Eq.~\eqref{eq:firstvariation} must hold for arbitrary variations of these quantities, the following set of microscale balance equations results
\begin{equation}
\begin{aligned}
\delta\vec{w}&:\ \left\{
\begin{aligned}
\vec{\nabla}_\mathrm{m}\cdot\bs{P}^\mathsf{T} &= \vec{\mu} + \sum_{i=1}^{n}\nu_i\vec{\varphi}_i + \sum_{i = 1}^{n}\vec{\eta}_i \cdot (\vec{\varphi}_i\vec{X}_\mathrm{m}), \ &&\mbox{in}\ \Omega_\mathrm{m}, \\
\bs{P}\cdot\vec{N}_\mathrm{m} &= \pm\vec{\lambda}, \ &&\mbox{on}\ \partial\Omega_\mathrm{m}^\pm,
\end{aligned}\right.\\
\delta\vec{\lambda}&:\ \mbox{periodicity constraint for~$\vec{w}$, Eq.~\eqref{eq:periodicity}},\\
\delta\vec{\mu}, \delta\underline{\nu}, \delta\underline{\eta}&:\ \mbox{kinematic constraints for~$\vec{w}$, Eqs.~\eqref{eq:constraintsa}--\eqref{eq:constraintsc}}.\\
\end{aligned}
\label{eq:mgoverning}
\end{equation}
As a consequence of the constraint term introduced in Eq.~\eqref{eq:constraint_definition}, the right hand side of the first governing Eq.~\eqref{eq:mgoverning} involves Lagrange multipliers acting as body forces inside each RVE, anti-periodic condition for RVE boundary tractions, and an additional set of orthogonality constraints. These terms were not present in the original formulation, where the constraint equations were enforced differently. Although the body forces~$\vec{\mu}$, $v_i\vec{\varphi}_i$, and~$\vec{\eta}_i \cdot (\vec{\varphi}_i\vec{X}_\mathrm{m})$, are directly associated with the orthogonality constrains of Eqs.~\eqref{eq:constraintsa}--\eqref{eq:constraintsc} in the form of Lagrange multipliers, they can be introduced also directly at the level of governing equations, see e.g.~\cite[][Eqs.~(8) and~(9)]{Yvonnet2020}. At the macroscale, only the slowly varying fields $\vec{v}_0(\vec{X})$ and $v_i(\vec{X})$ are relevant, and their governing equations read
\begin{equation}
\delta\vec{v}_0:\ \left\{
\begin{aligned}
\vec{\nabla}\cdot\bs{\Theta}^\mathsf{T} &= \vec{0}, \quad \mbox{in}\ \Omega,\\
\bs{\Theta}\cdot\vec{N} &= \vec{0}, \quad \mbox{on}\ \Gamma_\mathrm{N}, \\
\end{aligned}\right.
\quad\delta v_i:\ \left\{
\begin{aligned}
\vec{\nabla}\cdot \vec{\Lambda}_{i}-\Pi_{i} &= 0, \quad \mbox{in}\ \Omega,\\
\vec{\Lambda}_{i}\cdot\vec{N} &= 0, \quad \mbox{on}\ \Gamma_\mathrm{N}, \\
\end{aligned}\right. \ i = 1, \dots, n,
\label{eq:Mbalanceequations}
\end{equation}
with the following definitions of macroscopic stress-like quantities:
\begin{align}
\bs{\Theta}(\vec{X}) &=  \frac{1}{|\Omega_\mathrm{m}|} \big\langle  \bs{P}(\vec{X}_\mathrm{m},\bs{F}) \, \big\rangle_{\Omega_\mathrm{m}}, 
\label{eq:Mstressesa}\\
\Pi_i(\vec{X}) &= \frac{1}{|\Omega_\mathrm{m}|} \big\langle \bs{P}(\vec{X}_\mathrm{m},\bs{F}):\vec{\nabla}_\mathrm{m}\vec{\varphi}_i(\vec{X}_\mathrm{m}) \, \big\rangle_{\Omega_\mathrm{m}}, 
\label{eq:Mstressesb}\\
\vec{\Lambda}_i(\vec{X}) &= \frac{1}{|\Omega_\mathrm{m}|} \big\langle \bs{P}^\mathsf{T}(\vec{X}_\mathrm{m},\bs{F})\cdot\vec{\varphi}_i(\vec{X}_\mathrm{m}) + \vec{X}_\mathrm{m}[\bs{P}(\vec{X}_\mathrm{m},\bs{F}):\vec{\nabla}_\mathrm{m}\vec{\varphi}_i(\vec{X}_\mathrm{m})] \, \big\rangle_{\Omega_\mathrm{m}}.\label{eq:Mstressesc}
\end{align}
The dependence of the homogenized stress quantities on the macroscopic position~$\vec{X}$ originates from the dependence of the deformation gradient~$\bs{F}(\vec{u}(\vec{X},\vec{X}_\mathrm{m}))$ on the macroscopic position through all smooth fields~$\vec{v}_0(\vec{X})$ and~$v_i(\vec{X})$, cf. also Eq.~\eqref{eq:approxu}.
%
%
\subsection{Second Variation}
The second variation of the Lagrangian~$\mathcal{L}$ reads
\begin{equation}
\begin{aligned}
\delta^2 \mathcal{L}(\vec{u},\vec{\mu}&,\underline{\nu},\underline{\eta},\vec{\lambda};\delta\vec{u},\delta\vec{\mu},\delta\underline{\nu},\delta\underline{\eta},\delta\vec{\lambda}) = \frac{1}{|\Omega_\mathrm{m}|} \Big\langle
\big\langle\vec{\nabla}_\mathrm{m}\delta\vec{u}: {^4}\bs{D}:\vec{\nabla}_\mathrm{m}\delta\vec{u}\big\rangle_{\Omega_\mathrm{m}} 
+ 2\, \delta\vec{\mu} \cdot \big\langle \delta\vec{w} \big\rangle_{\Omega_\mathrm{m}} \\
&+ 2\sum_{i=1}^{n} \delta\nu_i \, \big\langle \delta\vec{w} \cdot \vec{\varphi}_i \big\rangle_{\Omega_\mathrm{m}} 
+ 2\sum_{i=1}^{n} \delta\vec{\eta}_i \cdot \big\langle \delta\vec{w} \cdot [ \vec{\varphi}_i\vec{X}_\mathrm{m} ] \big\rangle_{\Omega_\mathrm{m}}
- 2 \, \big\langle \delta\vec{\lambda}\cdot \llbracket \delta\vec{w} \rrbracket \big\rangle_{\partial\Omega_\mathrm{m}^+}
\Big\rangle_{\Omega},
\end{aligned}
\label{eq:secondvariation}
\end{equation}
where the spatial dependencies on~$\vec{X}$ and~$\vec{X}_\mathrm{m}$ have been dropped for brevity, and where the local stiffness tensor~${{^4}\bs{D}}_\mathrm{m}(\vec{X}_\mathrm{m},\bs{F})$ has been introduced as
\begin{equation}
{^4}\bs{D}(\vec{X}_\mathrm{m},\bs{F}) = \frac{\partial^2 \Psi(\vec{X}_\mathrm{m},\bs{F}(\vec{u}(\vec{X},\vec{X}_\mathrm{m})))}{\partial\bs{F}^\mathsf{T}\partial\bs{F}^\mathsf{T}}.
\end{equation}
Positive definiteness of the modified Lagrangian, from which all constraint variables---i.e.~Lagrange multipliers---have been condensed out, reflects stability of the combined multiscale system including micro- as well as macro-quantities. Upon condensing out also the microfluctuation field~$\vec{w}$, stability of the macroscopic system can be assessed. The microfluctuation field as well as all Lagrange multipliers can be eliminated for each macroscopic point~$\vec{X}$ by static condensation. This is done by means of the Schur complement after the FE numerical discretization, as described in Section~\ref{ch:numerical} below, whereas stability of the macroscopic system is detailed in Section~\ref{ch:stability}.

Substituting Eq.~\eqref{eq:du} into the first expression on the right hand side of Eq.~\eqref{eq:secondvariation}, one obtains a set of~$ p(p+1)/2 $, $p = n+2$, coupled specific stiffness quantities for the macroscopic, $\vec{v}_0$, $v_i$, $i = 1,\dots,n$, and microscopic, $\vec{w}$, fields. In particular, the individual terms corresponding to the macroscopic quantities read as
\begingroup
\allowdisplaybreaks
\begin{align}
{^4}\bs{D}_{00} &= \big\langle{{^4}\bs{D}}\big\rangle_{\Omega_\mathrm{m}},\label{eq:Mstiffnessesa}\\
\bs{D}_{0i_\mathsf{N}}=\bs{D}_{i_\mathsf{N}0} &= \big\langle {{^4}\bs{D}}:\vec{\nabla}_\mathrm{m}\vec{\varphi}_i \big\rangle_{\Omega_\mathrm{m}}, 
\label{eq:Mstiffnessesb}\\
{{^3}\bs{D}}_{0i_\mathsf{B}} = \big({{^3}\bs{D}}^\mathsf{LT}_{i_\mathsf{B}0}\big)^\mathsf{RT} &= \big\langle {{^4}\bs{D}}^\mathsf{RT}\cdot\vec{\varphi}_i + [({^4}\bs{D}:\vec{\nabla}_\mathrm{m}\vec{\varphi}_i) \vec{X}_\mathrm{m}]  \big\rangle_{\Omega_\mathrm{m}},
\label{eq:Mstiffnessesc}\\
{D}_{i_\mathsf{N}j_\mathsf{N}} &= \big\langle \vec{\nabla}_\mathrm{m}\vec{\varphi}_i
:
{{^4}\bs{D}}:
\vec{\nabla}_\mathrm{m}\vec{\varphi}_j \big\rangle_{\Omega_\mathrm{m}}, 
\label{eq:Mstiffnessesd}\\
{\vec{D}}_{i_\mathsf{N}j_\mathsf{B}} = {\vec{D}}_{i_\mathsf{B}j_\mathsf{N}} &= \big\langle
\vec{\nabla}_\mathrm{m}\vec{\varphi}_i 
: [ \, {{^4}\bs{D}}^\mathsf{RT} \cdot\vec{\varphi}_j  
+ ({{^4}\bs{D}} :\vec{\nabla}_\mathrm{m}\vec{\varphi}_j) \vec{X}_\mathrm{m} \, ]\big\rangle_{\Omega_\mathrm{m}},
\label{eq:Mstiffnessese}\\
{\bs{D}}_{i_\mathsf{B}j_\mathsf{B}} &= \big\langle
\vec{\varphi}_i \cdot  {{^4}\bs{D}}^\mathsf{RT} \cdot \vec{\varphi}_j +
(\vec{\varphi}_i \cdot  {{^4}\bs{D}} : \vec{\nabla}_\mathrm{m}\vec{\varphi}_j) \vec{X}_\mathrm{m} \nonumber\\
&+ \vec{X}_\mathrm{m}(\vec{\nabla}_\mathrm{m}\vec{\varphi}_i : {{^4}\bs{D}}^\mathsf{RT} \cdot \vec{\varphi}_j) + \vec{X}_\mathrm{m}(\vec{\nabla}_\mathrm{m}\vec{\varphi}_i : {{^4}\bs{D}} : \vec{\nabla}_\mathrm{m}\vec{\varphi}_j) \vec{X}_\mathrm{m} \big\rangle_{\Omega_\mathrm{m}},\label{eq:Mstiffnessesg}
\end{align}
\endgroup
where the subscript~$0$ relates to~$\vec{\nabla}\delta\vec{v}_0$, $i_\mathsf{B}$ to~$\vec{\nabla}\delta{v}_i$, and~$i_\mathsf{N}$ to~$\delta{v}_i$ terms, respectively, with~$i = 1,\dots, n$, and~$j = 1,\dots,n$, which are essential for numerical solution and stability assessment of the macroscopic system.
%
%
\section{Numerical Implementation}
\label{ch:numerical}
At every macroscopic integration point, the macroscopic quantities~$\vec{\nabla}\vec{v}_0$, $v_i$, and~$\vec{\nabla} v_i$, are sampled and passed down to the microscale, cf.~Fig.~\ref{fig:comphomog}, where the modes~$\vec{\varphi}_i$ and microfluctuation field~$\vec{w}$ are defined. Taking into account the constraints~\eqref{eq:constraintsa}--\eqref{eq:periodicity}, the microfluctuation field~$\vec{w}$ is computed by solving the microscale boundary value problem defined in Eq.~\eqref{eq:mgoverning}. Knowing~$\vec{w}$, the homogenized macroscopic stresses and stiffnesses are computed following Eqs.~\eqref{eq:Mstressesa}--\eqref{eq:secondvariation} and~\eqref{eq:Mstiffnessesa}--\eqref{eq:Mstiffnessesg}. All of these quantities are required for the solution of the macroscopic boundary value problem of Eq.~\eqref{eq:Mbalanceequations} using the standard Newton algorithm, leading to a fast quadratic convergence and allowing for a bifurcation analysis. Although multiple approaches can be adopted for the solution of the multiscale problem, see e.g.~\citep{Okada:2010}, here we adopt the condensation method. The discretization and numerical solution of the microscopic problem is presented in Section~\ref{ch:microscopicproblem}. The macroscopic problem is subsequently elaborated upon in Section~\ref{ch:macroscopicproblem}. The bifurcation analysis is detailed in Section~\ref{ch:stability}, including a nested algorithmic scheme. A Matlab implementation of the presented MicroMorphic homogenization for Multiscale Metamaterials framework~(mm4mm) is available at GitLab's repository~\href{https://gitlab.com/rokosondrej/mm4mm.git}{mm4mm}.
%
%
\subsection{Microscopic Problem}
\label{ch:microscopicproblem}
Using standard FE procedures, the microfluctuation field~$\vec{w}$ and its gradient~$\vec{\nabla}\vec{w}$ are expressed over each microscopic element~$e$ within an RVE in terms of the shape functions and nodal values as
\begin{equation}
\begin{aligned}
\vec{w}(\vec{X},\vec{X}_\mathrm{m}) \approx \bs{\mathsf{N}}^e_w(\vec{X}_\mathrm{m})\underline{w}^e(\vec{X}), \quad
\vec{\nabla}_\mathrm{m}\vec{w}(\vec{X},\vec{X}_\mathrm{m}) \approx \bs{\mathsf{B}}^e_w(\vec{X}_\mathrm{m})\underline{w}^e(\vec{X}),
\end{aligned}
\label{eq:discretizationw}
\end{equation}
where~$\underline{w}^e$ is a column of element nodal values of the~$\vec{w}$ field, collected for all~$n_\mathrm{e}$ elements in a column matrix~$\column{w}$, $\bs{\mathsf{N}}_w^e$ is a matrix of the corresponding shape functions, and~$\bs{\mathsf{B}}_w^e$ a matrix of the shape function derivatives. Corresponding variations~$\delta\vec{w}$ and~$\vec{\nabla}\delta\vec{w}$ are discretized in the same way. The patterning modes are expressed similarly as
\begin{equation}
\vec{\varphi}_i(\vec{X}_\mathrm{m}) \approx \bs{\mathsf{N}}^e_w(\vec{X}_\mathrm{m}) \underline{\varphi}_i^e, \quad
\vec{\nabla}\vec{\varphi}_i(\vec{X}_\mathrm{m}) \approx \bs{\mathsf{B}}^e_w(\vec{X}_\mathrm{m}) \underline{\varphi}_i^e, \label{eq:discretizationphi}
\end{equation}
where~$\underline{\varphi}_i^e$ is a column storing element nodal values of~$\vec{\varphi}_i$, collected over all elements in a column matrix~$\column{\varphi}_i$. Although analytical expressions for the patterning modes~$\vec{\varphi}_i$ are provided below in Eqs.~\eqref{eq:modeSquare} and~\eqref{eq:mode_1}, we opted here for a discretized version for convenience and generality, since alternative definitions of patterning modes may be more appropriate, e.g.~based on linearized buckling analysis or true deformed shapes obtained numerically. The discretized version of the orthogonality constraint~\eqref{eq:constraintsa} takes the form
\begin{equation}
\big\langle \vec{w}(\vec{X},\vec{X}_\mathrm{m}) \big\rangle_{\Omega_\mathrm{m}} 
\approx
\left\{
\begin{aligned}
\underline{1}_1^\mathsf{T} \bs{\mathsf{M}} \underline{w} &= 0, \\
\underline{1}_2^\mathsf{T} \bs{\mathsf{M}} \underline{w} &= 0,
\end{aligned}
\right.
\label{eq:discretizedRbm}
\end{equation}
where~$\bs{\mathsf{M}} = \Aop_{e=1}^{n_\mathrm{e}}\langle \bs{(\mathsf{N}}^e_w)^\mathsf{T} \bs{\mathsf{N}}^e_w \rangle_{\Omega_\mathrm{m}^e}$ is the symmetric Gramian matrix of microscopic shape functions, and~$\underline{1}_k$ is a column matrix with ones at the positions of Degrees Of Freedom~(DOFs) corresponding to the $k$-th component (i.e.~either horizontal or vertical component) of~$\vec{w}$. $\Aop$ denotes the standard FE assembly operator, and the integration over each microcopic element volume~$\Omega_\mathrm{m}^e$ is performed using a standard Gauss integration rule, for brevity not expressed explicitly as a sum. Analogously, the discrete forms of the scalar constraints~\eqref{eq:constraintsb} read
\begin{equation}
\big\langle \vec{w}(\vec{X},\vec{X}_\mathrm{m}) \cdot \vec{\varphi}_i(\vec{X}_\mathrm{m}) \big\rangle_{\Omega_\mathrm{m}} \approx
\underline{\varphi}_i^\mathsf{T} \bs{\mathsf{M}} \underline{w} = 0, \quad i = 1, \dots, n,
\label{eq:discretizedPhi}
\end{equation}
whereas the set of vector constraints~\eqref{eq:constraintsc} is discretized as
\begin{equation}
\big\langle \vec{w}(\vec{X},\vec{X}_\mathrm{m}) \cdot [ \vec{\varphi}_i(\vec{X}_\mathrm{m}) \vec{X}_\mathrm{m} ] \big\rangle_{\Omega_\mathrm{m}} 
\approx
\left\{
\begin{aligned}
\underline{q}_{i,1}^\mathsf{T} \bs{\mathsf{M}} \underline{w} &= 0, \\
\underline{q}_{i,2}^\mathsf{T} \bs{\mathsf{M}} \underline{w} &= 0, 
\end{aligned}
\right. \quad i = 1,\dots,n,
\label{eq:discretizedGradPhi}
\end{equation}
where~$\underline{q}_{i,k}$ stores~$(\vec{\varphi}_i\cdot\vec{e}_1) X_k$ and~$(\vec{\varphi}_i\cdot\vec{e}_2) X_k$ at the positions of the DOFs corresponding to the first and second component. The periodicity constraint of Eq.~\eqref{eq:periodicity} is expressed with the help of the link topology matrix~$\bs{\mathsf{C}}_\mathrm{pbc}$, described e.g.~in~\citep{Miehe:2007}, as
\begin{equation}
\llbracket \vec{w}(\vec{X},\vec{X}_\mathrm{m}) \rrbracket \approx \bs{\mathsf{C}}_\mathrm{pbc} \underline{w} = \underline{0}.
\label{eq:discretisedperiodicity}
\end{equation}
Finally, the microscopic governing equation~\eqref{eq:mgoverning} is solved iteratively using the standard Newton method~\citep[see][Section~14]{Bonnans:2006} for the linear system
\begin{equation}
\begin{bmatrix}
	\bs{\mathsf{D}}_{ww} & \bs{\mathsf{C}}_\mathrm{pbc}^\mathsf{T} & \bs{\mathsf{M}} \underline{1}_1 &\dotsc& \bs{\mathsf{M}} \underline{q}_{n,2} \\
	\bs{\mathsf{C}}_\mathrm{pbc} & 0 & \multicolumn{2}{c}{\smash{\dotsc}} & {0}\\
	\underline{1}_1^\mathsf{T} \bs{\mathsf{M}}& {\smash{\raisebox{-0.8\normalbaselineskip}{$\bs{\vdots}$}}}& \multicolumn{2}{c}{\smash{\raisebox{-0.8\normalbaselineskip}{$\ddots$}}}&{\smash{\raisebox{-0.8\normalbaselineskip}{$\bs{\vdots}$}}}\\
	\vdots\\
	\underline{q}_{n,2}^\mathsf{T}\bs{\mathsf{M}} & 0 & \multicolumn{2}{c}{\smash{\dotsc}}& {0}
\end{bmatrix}
\begin{bmatrix}
	d\underline{w}\\
	\underline{\lambda}\\
	\underline{\mu}\\
	\underline{\nu}\\
	\underline{\eta}
\end{bmatrix} = -
\begin{bmatrix}
	\underline{f}_w\\
	\underline{0}\\
	\underline{0}\\
	\underline{0}\\
	\underline{0}
\end{bmatrix}, 
\label{eq:microscalesolving}
\end{equation}
or in a compact form as
\begin{equation}
\bs{\mathsf{D}}_{ww}^\star d\underline{w}^\star = -\underline{f}_w^\star.
\label{eq:microNewton}
\end{equation}
In Eqs.~\eqref{eq:microscalesolving} and~\eqref{eq:microNewton}, 
\begin{equation}
\bs{\mathsf{D}}_{ww} = {\textstyle\Aop_{e=1}^{n_\mathrm{e}}} \big\langle
(\bs{\mathsf{B}}^e_w)^\mathsf{T} \,
{\bs{\mathsf{D}}}
\bs{\mathsf{B}}^e_w\big\rangle_{\Omega^e_\mathrm{m}},
\label{eq:mdiscretisedmstiffnessesd}
\end{equation}
is the microscopic stiffness matrix, $\underline{f}_w^\star$ is the column of microscopic nodal internal forces~$\column{f}_w = \Aop_{e=1}^{n_\mathrm{e}} \langle (\mtrx{B}_w^e)^\mathsf{T}\column{P} \rangle_{\Omega_\mathrm{m}^e}$ and unbalanced equality constraints, where~$\column{P}$ is a column storing the components of the stress tensor~$\tensor{P}$, and $d\underline{w}^\star$ is a column storing the iterative change of the microscopic fluctuation field and current iterative values of Lagrange multipliers. Note that, in analogy to~$\column{\eta}$ and~$\column{\nu}$ in Eq.~\eqref{micromorphic:eq7}, $\column{\mu}$ stores individual components of~$\vec{\mu}$. An asterisk indicates that the matrix accounts for all the equality constraints and thus also has the corresponding extra entries. To evaluate the vector of current internal forces~$\underline{f}_w$ and stiffness matrix~$\mtrx{D}_{ww}$, the nodal values of the entire displacement field~$\column{u}$ of Eq.~\eqref{eq:approxu} need to be constructed over the entire RVE from the knowledge of the current state of the macroscopic quantities~$\vec{\nabla}\vec{v}_0$, $v_i$, $\vec{\nabla}v_i$, patterning modes~$\vec{\varphi}_i$, $i = 1,\dots, n$, and iterative state of the microfluctuation field~$\underline{w}$. Because the primary buckling modes are captured by the patterning fields~$\vec{\varphi}_i$, a microscopic bifurcation analysis and stability control analogous to Section~\ref{ch:stability} below is not required.
%
%
\subsection{Macroscopic Problem}
\label{ch:macroscopicproblem}
The macroscopic fields $\vec{v}_0$ and $v_i$ are discretized within each macroscopic element~$E$ as
\begin{equation}
\begin{aligned}
\vec{v}_0(\vec{X}) &\approx \bs{\mathsf{N}}^E_0(\vec{X})\underline{v}_0^E, \quad & v_i(\vec{X}) &\approx  \bs{\mathsf{N}}^E_i(\vec{X})\underline{v}_i^E, \\
\vec{\nabla}\vec{v}_0(\vec{X}) &\approx \bs{\mathsf{B}}^E_0(\vec{X})\underline{v}_0^E, \quad 
&\vec{\nabla}v_i(\vec{X}) &\approx  \bs{\mathsf{B}}^E_i(\vec{X})\underline{v}_i^E,
\end{aligned}
\label{simplified:discretization:eq111}
\end{equation}
where~$\mtrx{N}^E_\bullet$ and~$\mtrx{B}^E_\bullet$ are macroscopic element shape functions and~$\column{v}_\bullet^E$ the corresponding column matrices of DOFs, collected globally for all~$n_\mathrm{E}$ elements in column matrices~$\column{v}_0$ and~$\column{v}_i$; the same forms and expressions are used to discretize their variations. The internal element forces
\begin{equation} 
\underline{f}_\mathrm{M}^E
= \begin{bmatrix}
\underline{f}_0^E\\
\underline{f}_1^E\\
\vdots\\
\underline{f}_n^E
\end{bmatrix}
\label{eq:fMig}
\end{equation}
are then obtained as a sum over~$n_\mathrm{g}$ macroscopic Gauss quadrature points, expressed explicitly as
\begin{align}
\column{f}_0^E &= \sum_{i_\mathrm{g}=1}^{n_\mathrm{g}}\column{f}_0^{E,i_\mathrm{g}} = \sum_{i_\mathrm{g}=1}^{n_\mathrm{g}}w_{i_\mathrm{g}}J_{i_\mathrm{g}} \, (\mtrx{B}_0^E)^\mathsf{T}\column{\Theta}^{i_\mathrm{g}}, \label{eq:elemForcesa}\\
\column{f}_i^E &= \sum_{i_\mathrm{g}=1}^{n_\mathrm{g}}\column{f}_i^{E,i_\mathrm{g}} =  \sum_{i_\mathrm{g}=1}^{n_\mathrm{g}}w_{i_\mathrm{g}}J_{i_\mathrm{g}}\big\{(\mtrx{B}_i^E)^\mathsf{T}\column{\Lambda}_i^{i_\mathrm{g}} - (\mtrx{N}_i^E)^\mathsf{T}\Pi_i^{i_\mathrm{g}}\big\}, \quad i = 1,\dots,n,
\label{eq:elemForces}
\end{align}
and assembled over all macroscopic elements to form the global internal force vector~$ \underline{f}_\mathrm{M} = \Aop_{E=1}^{n_\mathrm{E}} \underline{f}_\mathrm{M}^E $. In Eqs.~\eqref{eq:elemForcesa} and~\eqref{eq:elemForces}, $w_{i_\mathrm{g}}$ are integration weights with the corresponding Jacobians~$J_{i_\mathrm{g}}$, and the column matrices~$\column{\Theta}^{i_\mathrm{g}}$, $\column{\Lambda}_i^{i_\mathrm{g}}$, and scalars~$\Pi_i^{i_\mathrm{g}}$, store the components of the homogenized stress quantities, defined in Eqs.~\eqref{eq:Mstressesa}--\eqref{eq:Mstressesc}, evaluated at appropriate positions of associated integration points.

In order to condense out the effect of the constrained microfluctuation field~$\column{w}$ and all Lagrange multipliers, the following monolithic incremental system of equations is assembled at each macroscopic quadrature point~$i_\mathrm{g}$ of each element~$E$, including both macroscopic as well as microscopic quantities (superscripts~$E$ and~$i_\mathrm{g}$ are dropped in Eqs.~\eqref{eq:Kuf}--\eqref{eq:mdiscretisedmstiffnessesc} for brevity):
\begin{equation}
\begin{bmatrix}
\bs{\mathsf{K}}_{00} & \bs{\mathsf{K}}_{01} & \dotsm & \bs{\mathsf{K}}_{0n} &\bs{\mathsf{K}}_{0w}^\star \\
\bs{\mathsf{K}}_{10} & \bs{\mathsf{K}}_{11} & \cdots & \bs{\mathsf{K}}_{1n} & \bs{\mathsf{K}}_{1w}^\star \\
\vdots & \vdots & \ddots & \vdots & \vdots\\
\bs{\mathsf{K}}_{n0} & \bs{\mathsf{K}}_{n1} & \cdots & \bs{\mathsf{K}}_{nn} & \bs{\mathsf{K}}_{nw}^\star \\
\bs{\mathsf{K}}_{w0}^\star & \bs{\mathsf{K}}_{w1}^\star & \cdots & \bs{\mathsf{K}}_{wn}^\star & \bs{\mathsf{K}}_{ww}^\star 
\end{bmatrix}
\begin{bmatrix}
\underline{v}_0^E\\
\underline{v}_1^E\\
\vdots\\
\underline{v}_n^E\\
\underline{w}^\star
\end{bmatrix} +
\begin{bmatrix}
\underline{f}_0\\
\underline{f}_1\\
\vdots\\
\underline{f}_n\\
{\underline{f}}_w^\star
\end{bmatrix}
= 
\begin{bmatrix}
\underline{r}_0\\
\underline{r}_1\\
\vdots\\
\underline{r}_n\\
\underline{0}
\end{bmatrix}.
\label{eq:Kuf}
\end{equation}
In Eq.~\eqref{eq:Kuf}, $\column{r}_i$, $i = 0,\dots,n$, are the residuals reflecting the fact that equilibrium is not satisfied at the level of an integration point, but only for the entire assembly over all quadrature points of all elements. The last row is zero, since the microscopic system of Eq.~\eqref{eq:microNewton} has been equilibrated, implying also that~${\underline{f}}_w^\star = \column{0}$. The specific stiffness for the macroscopic quantities can be then obtained as a Schur complement via static condensation:
\begin{equation} 
\bs{\mathsf{K}}^{E,i_\mathrm{g}}_\mathrm{M}
= 
\begin{bmatrix}
\bs{\mathsf{K}}_{00} & \bs{\mathsf{K}}_{01} & \dotsm & \bs{\mathsf{K}}_{0n} \\
\bs{\mathsf{K}}_{10} & \bs{\mathsf{K}}_{11} & \cdots & \bs{\mathsf{K}}_{1n} \\
\vdots & \vdots & \ddots & \vdots \\
\bs{\mathsf{K}}_{n0} & \bs{\mathsf{K}}_{n1} & \cdots & \bs{\mathsf{K}}_{nn} 
\end{bmatrix}
-
\begin{bmatrix}	
\bs{\mathsf{K}}_{0w}^\star \\ 
\bs{\mathsf{K}}_{1w}^\star \\
\vdots \\
\bs{\mathsf{K}}_{nw}^\star
\end{bmatrix}
\begin{bmatrix}
\bs{\mathsf{K}}_{ww}^\star
\end{bmatrix}^{-1}
\begin{bmatrix}
\bs{\mathsf{K}}_{w0}^\star & \bs{\mathsf{K}}_{w1}^\star & \dotsm & \bs{\mathsf{K}}_{wn}^\star
\end{bmatrix}.
\label{eq:condensationK}
\end{equation}
Note that whereas for a given element~$E$ an~$n_\mathrm{g}$ microfluctuation fields~$\column{w}^{\star,E,i_\mathrm{g}}$ are computed and condensed out, only one set of macroscopic DOFs for the coarse fields~$\column{v}_0^E$ and~$\column{v}_i^E$ pertinent to that element are involved. The asterisk superscript to the specific stiffness components in Eqs.~\eqref{eq:Kuf} and~\eqref{eq:condensationK} again indicates that these matrices have extra zero entries corresponding to the Lagrange multipliers. The individual sub-matrices are defined as
\begingroup
\allowdisplaybreaks
\begin{align}
\bs{\mathsf{K}}_{00} &= {\frac{1}{|\Omega_\mathrm{m}|}} \bs{\mathsf{B}}_0^\mathsf{T} \bs{\mathsf{D}}_{00} \bs{\mathsf{B}}_0, \label{eq:Mdiscretisedstiffnessesa}\\
\bs{\mathsf{K}}_{0i} = \bs{\mathsf{K}}_{i0}^\mathsf{T} &= {\frac{1}{|\Omega_\mathrm{m}|}} \big( \bs{\mathsf{B}}_0^\mathsf{T} \bs{\mathsf{D}}_{0i_\mathsf{B}} \bs{\mathsf{B}}_i  + \bs{\mathsf{B}}_0^\mathsf{T} \underline{\mathsf{D}}_{0i_\mathsf{N}} \bs{\mathsf{N}}_i \big), \label{eq:Mdiscretisedstiffnessesb}\\
\bs{\mathsf{K}}_{ij} &= {\frac{1}{|\Omega_\mathrm{m}|}} \big( \bs{\mathsf{N}}_i^\mathsf{T} \mathsf{D}_{i_\mathsf{N}j_\mathsf{N}} \bs{\mathsf{N}}_j + \bs{\mathsf{B}}_i^\mathsf{T} \underline{\mathsf{D}}_{i_\mathsf{B}j_\mathsf{N}} \bs{\mathsf{N}}_j + \bs{\mathsf{N}}_i^\mathsf{T} \underline{\mathsf{D}}_{i_\mathsf{B}j_\mathsf{N}}^\mathsf{T} \bs{\mathsf{B}}_j + \bs{\mathsf{B}}_i^\mathsf{T} \bs{\mathsf{D}}_{i_\mathsf{B}j_\mathsf{B}} \bs{\mathsf{B}}_j \big), \label{eq:Mdiscretisedstiffnessesc}\\
\bs{\mathsf{K}}_{0w} = \bs{\mathsf{K}}_{w0}^\mathsf{T} &= {\frac{1}{|\Omega_\mathrm{m}|}}  \bs{\mathsf{B}}_0^\mathsf{T} \bs{\mathsf{D}}_{0w}, \label{eq:Mdiscretisedstiffnessesd}\\
\bs{\mathsf{K}}_{iw} = \bs{\mathsf{K}}_{wi}^\mathsf{T} &= {\frac{1}{|\Omega_\mathrm{m}|}} \big( \bs{\mathsf{N}}_i^\mathsf{T} \underline{\mathsf{D}}_{i_\mathsf{N}w} +
\bs{\mathsf{B}}_i^\mathsf{T} \bs{\mathsf{D}}_{i_\mathsf{B}w} \big)   ,\label{eq:Mdiscretisedstiffnessese}\\
\bs{\mathsf{K}}_{ww}^\star &=  {\frac{1}{|\Omega_\mathrm{m}|}} \bs{\mathsf{D}}_{ww}^\star,\label{eq:Mdiscretisedstiffnessesf}
\end{align}
\endgroup
where~${\bs{\mathsf{D}}}_{00}$, $\underline{\mathsf{D}}_{0i_\mathsf{N}} = \underline{\mathsf{D}}_{i_\mathsf{N}0}^\mathsf{T}$, $\bs{\mathsf{D}}_{0i_\mathsf{B}} = {\bs{\mathsf{D}}}_{i_\mathsf{B}0}^\mathsf{T}$, ${\mathsf{D}}_{i_\mathsf{N}j_\mathsf{N}}$, $\underline{\mathsf{D}}_{i_\mathsf{N}j_\mathsf{B}}$, and~${\bs{\mathsf{D}}}_{i_\mathsf{B}j_\mathsf{B}}$, are matrix representations of the specific stiffnesses defined in Eqs.~\eqref{eq:Mstiffnessesa}--\eqref{eq:Mstiffnessesg}, obtained upon RVE discretization simply as volume integrals using a standard Gauss integration rule. For instance, ${\bs{\mathsf{D}}}$ is a $4\times4$ matrix representation of the microscopic fourth-order stiffness tensor~${^4}\bs{D}$, etc. In addition to the expressions~\eqref{eq:Mdiscretisedstiffnessesa}--\eqref{eq:Mdiscretisedstiffnessesf}, there are $ 3n + 1 $ coupled specific stiffnesses related to the microscopic variation~$\vec{\nabla}\delta\vec{w}$ and one of the macroscopic variations~$\vec{\nabla}\delta\vec{v}_0$, $\delta{v}_i$, or $\vec{\nabla}\delta{v}_i$, and one microscopic stiffness quantity that relates twice to~$\vec{\nabla}\delta\vec{w}$. Because~$\vec{\nabla}\delta\vec{w}$ depends on the microscopic position~$\vec{X}_\mathrm{m}$, over which the integral in Eq.~\eqref{eq:secondvariation} is carried out, these specific stiffnesses can be written only upon RVE discretization, yielding
\begingroup
\allowdisplaybreaks
\begin{align}
\bs{\mathsf{D}}_{0w} = \bs{\mathsf{D}}^\mathsf{T}_{w0} &= {\textstyle\Aop_{e=1}^{n_\mathrm{e}}} \big\langle {\bs{\mathsf{D}}} \bs{\mathsf{B}}^e_w \big\rangle_{\Omega^e_\mathrm{m}},
\label{eq:mdiscretisedmstiffnessesa}\\
\underline{\mathsf{D}}_{i_\mathsf{N}w} = \underline{\mathsf{D}}^\mathsf{T}_{wi_\mathsf{N}} &= {\textstyle\Aop_{e=1}^{n_\mathrm{e}}} \big\langle
(\bs{\mathsf{B}}^e_w\underline{\varphi}^e_i)^\mathsf{T}
{\bs{\mathsf{D}}}
\bs{\mathsf{B}}^e_w \big\rangle_{\Omega^e_\mathrm{m}},\label{eq:mdiscretisedmstiffnessesb}
\\
\bs{\mathsf{D}}_{i_\mathsf{B}w} = \bs{\mathsf{D}}^\mathsf{T}_{wi_\mathsf{B}} &= {\textstyle\Aop_{e=1}^{n_\mathrm{e}}} \big\langle
\big\{
(\bs{\mathsf{N}}^e_w\underline{\varphi}^e_i)^\mathsf{T}
{\bs{\mathsf{D}}} 
+ \underline{X}_\mathrm{m} \, (\bs{\mathsf{B}}^e_w\underline{\varphi}^e_i)^\mathsf{T}
{\bs{\mathsf{D}}}  \big\}
\bs{\mathsf{B}}^e_w\, \big\rangle_{\Omega^e_\mathrm{m}},\label{eq:mdiscretisedmstiffnessesc}
\end{align}
\endgroup
where the integration over~$\Omega_\mathrm{m}^e$ is again carried out numerically. 

The stiffness matrix of the entire macroscopic element is obtained by summing the contributions from all quadrature points,
\begin{align}
\mtrx{K}_\mathrm{M}^E &= \sum_{i_\mathrm{g}=1}^{n_\mathrm{g}}w_{i_\mathrm{g}}J_{i_\mathrm{g}}\mtrx{K}_\mathrm{M}^{E,i_\mathrm{g}},
\label{eq:elemStiffness}
\end{align}
which are eventually assembled into a global stiffness matrix~$ \mtrx{K}_\mathrm{M} = \Aop_{E=1}^{n_\mathrm{E}} \mtrx{K}_\mathrm{M}^E $. The resulting macroscopic system is again solved using the standard Newton method with an incremental system of linear equations
\begin{equation}
\bs{\mathsf{K}}_\mathrm{M} d\underline{v}_\mathrm{M} = \column{f}_\mathrm{ext}-\underline{f}_\mathrm{M},
\end{equation}
where~$\column{f}_\mathrm{ext}$ denotes a column of externally applied forces (acting only on~$\column{v}_0$), and
\begin{equation}
d\underline{v}_\mathrm{M} = 
\begin{bmatrix}
d\underline{v}_0\\
d\underline{v}_1\\
\vdots\\
d\underline{v}_n
\end{bmatrix}
\end{equation}
is an iterative increment of the global macroscopic quantities.
%
%
\subsection{Bifurcation analysis} 
\label{ch:stability}
Following~\cite{Miehe:2002}, an equilibrated configuration~$\underline{v}_\mathrm{M}$ of a system is considered to be stable if the energy of this state is lower than the energy associated with a state obtained by adding a small kinematically admissible perturbation~$\delta\underline{v}_\mathrm{M}$ to the equilibrated configuration. That is, if
\begin{equation}
\widehat{\mathcal{E}}(\underline{v}_\mathrm{M} + \delta\underline{v}_\mathrm{M}) - \widehat{\mathcal{E}}(\underline{v}_\mathrm{M}) \approx \frac{1}{2} \delta\underline{v}_\mathrm{M}^\mathsf{T}\bs{\mathsf{K}}_{\mathrm{M}}(\underline{v}_\mathrm{M})\delta\underline{v}_\mathrm{M} > 0,
\label{eq:Mstability}
\end{equation}
where the second-order Taylor series expansion of the total energy has been used. Notice that~$\widehat{\mathcal{E}}$ corresponds to the total potential energy of the entire system (Eq.~\eqref{eq:energy_definition}) from which microfluctuation fields and Lagrange multipliers, $\column{w}^\star$, associated with all macroscopic Gauss points have been condensed out, and that the first-order term vanishes as a result of equilibrium. The condition~\eqref{eq:Mstability} is equivalent to the requirement of positive definiteness of~$\mtrx{K}_\mathrm{M}$, i.e.~to the requirement that all eigenvalues of~$\mtrx{K}_\mathrm{M}$ are positive. If the lowest eigenvalue~$\alpha_1$ is non-positive, the associated configuration is unstable and the corresponding eigenvector~$\underline{\psi}_1$ determines the buckling mode. The equilibrated solution~$\underline{v}_\mathrm{M}$ of the current increment is then perturbed with the eigenvector~$\underline{\psi}_1$ multiplied by a small perturbation factor~$\tau > 0$,
\begin{equation}
\underline{v}_\mathrm{M} = \underline{v}_\mathrm{M} + \tau\underline{\psi}_1,
\end{equation}
and the system is equilibrated again. The factor~$\tau$ is increased until a stable equilibrium is reached, i.e.~until a possible energy barrier between the current unstable and a stable buckled configuration is overcome, and at the same time until the lowest eigenvalue~$\alpha_1$ of the updated macroscopic stiffness matrix~$\mtrx{K}_\mathrm{M}$ does not become positive. If the system fails to find a stable equilibrium even for large~$\tau$, the previous increment is halved to decrease the energy barrier, and the entire procedure is repeated.

An outline of the overall micromorphic computational homogenization scheme for multiple modes, including stability control, is given in Algorithm~\ref{algorithm}.
\begin{algorithm}[htbp]
	\begin{minipage}{\linewidth}
		\caption{Nested solution scheme for the micromorphic computational homogenization framework with a full Newton implementation and stability control.}
		\label{algorithm}
		\centering
		\vspace{-\topsep}
		\begin{enumerate}[1:]
			\item \textbf{Initialization:}
			\begin{enumerate}[(a):]
				\item Initialize the macroscopic model, $\column{v}_0(t = 0) = \column{0}$, $\column{v}_i(t=0) = \column{0}$ for all $i = 1, \dots, n$.
				\item Assign an RVE to each Gauss integration point of the macro-model.
			\end{enumerate}
			\item \textbf{for~$k=1,\dots,n_T$} (loop over all time steps of associated parametrization time)
			\begin{enumerate}[(i):]
				\item Apply macroscopic boundary conditions at time step~$k$.
				\item \textbf{while~$\epsilon > \mathrm{tol}$} (macroscopic solver, iteration~$l$)
				\begin{enumerate}[(a):]
					\item From~$\column{v}_0^{\,l}$ and~$\column{v}_i^l$ compute deformation gradient~$\bs{I}+(\vec{\nabla}\vec{v}_0^{\,i_\mathrm{g}})^\mathsf{T}$, mode magnitude~$v_i^{i_\mathrm{g}}$, and its gradient~$\vec{\nabla}v_i^{i_\mathrm{g}}$ for each macroscopic Gauss point~$i_\mathrm{g}$.
					\item Perform the RVE analysis for each macroscopic Gauss point~$i_\mathrm{g}$:
					\begin{itemize}[\textbf{-}]
						\item Apply underlying deformation dictated by~$\bs{I}+(\vec{\nabla}\vec{v}_0^{\,i_\mathrm{g}})^\mathsf{T}$, $v_i^{i_\mathrm{g}}$, $\vec{\nabla}v_i^{i_\mathrm{g}}$, and~$\column{\varphi}_i$.
						\item Assemble and solve the nonlinear RVE problem, Eq.~\eqref{eq:mgoverning}. For $\column{w}$ enforce orthogonality, periodicity, and rigid body motion constraints~\eqref{eq:constraintsa}--\eqref{eq:periodicity} over~$\Omega_\mathrm{m}$.
						\item Average resulting microscopic quantities to obtain the homogenized macroscopic stresses and stiffnesses.
					\end{itemize}
					\item Assemble the macroscopic gradient $\underline{f}^l_\mathrm{M}$ and tangent $\bs{\mathsf{K}}^l_\mathrm{M}$ by condensing out the stiffness terms related to $\column{w}$ and all Lagrange multipliers.
					\item Update the macroscopic displacements~$\underline{v}^{l+1}_\mathrm{M} = \underline{v}^{l}_\mathrm{M} + d\underline{v}^{l}_\mathrm{M}$, where $ \bs{\mathsf{K}}^l_\mathrm{M} d\underline{v}^{l}_\mathrm{M} = \column{f}_\mathrm{ext} -\underline{f}^l_\mathrm{M} $.
					\item Update the iteration error~$\epsilon = \|\underline{f}^l_\mathrm{M}\| + \|d\underline{v}^l_\mathrm{M}\|$.
				\end{enumerate}	
				\item \textbf{end while}
					\item If the lowest eigenvalue~$\alpha_1$ of~$\bs{\mathsf{K}}^l_\mathrm{M}$ is non-positive, perturb the system with corresponding eigenvector~$\tau\underline{\psi}_1$, and equilibrate iteratively for an increasing perturbation factor~$\tau$ until the system becomes stable. Then continue to~(i) for~$k = k+1$. If perturbation fails, halve the load increment and proceed to~(i) with current~$k$.
			\end{enumerate}
			\item \textbf{end for}
		\end{enumerate}
		\vspace{-\topsep}
	\end{minipage}
\end{algorithm}
%
%
\section{Numerical Examples}
\label{results}
In this section, predictions made using the micromorphic computational homogenization scheme, introduced in Sections~\ref{ch:micromorphic} and \ref{ch:numerical}, are compared against Direct Numerical Simulations~(DNS) for two examples. The first example represents a metamaterial column composed of a square stacking of holes subjected to compression, whereas the second example considers a specimen with a hexagonal stacking of holes subjected to a uniform compressive loading with various biaxiality ratios, buckling locally into one of multiple possible patterns.

The constitutive behaviour of the elastomer base material is modelled by a hyperelastic law with the following energy density
\begin{equation}
\psi( \bs{F} (\vec{X},\vec{X}_\mathrm{m}) ) = c_1 (I_1 - 3) + c_2 (I_1 - 3)^2 - 2 c_1 \log{J} + \frac{1}{2} K (J-1)^2,
\label{eq:energydensity}  
\end{equation}
where~$\bs{F}=\bs{I}+(\vec{\nabla}\vec{u})^\mathsf{T}
$ is the deformation gradient tensor, where the gradient operator~$\vec{\nabla}$ is defined with respect to the reference configuration, $J = \det{\bs{F}}$, and~$I_1 = \tr{\bs{C}}$ is the first invariant of the right Cauchy--Green deformation tensor~$\bs{C} = \bs{F}^\mathsf{T}\cdot\bs{F}$. The values of the constitutive parameters employed, listed in Tab.~\ref{table}, are based on the experimental characterization of~\cite{Bertoldi:2008}.

The smallest RVE domains of the size~$2\ell$ are adopted in both examples for the micromorphic homogenization scheme, as shown in Figs.~\ref{fig:columnMacroTriangulation} and~\ref{fig:hexab} below, which are large enough to accommodate the longest microstructural patterning modes (see e.g.~\citealt{Bertoldi:2008} for the square and~\citealt{Ohno:2002} for hexagonal stacking of holes). Although the chosen RVE size is sufficiently large to accommodate microstructural buckling, because of the periodicity assumption on the microfluctuation fields~$\vec{w}$ (recall Eq.~\eqref{eq:periodicity} and the discussion therein), choosing larger RVE domains may still slightly affect obtained results, especially for a vanishing separation of scales.
\begin{table}
	\centering
	\caption{Constitutive parameters of the hyperelastic law specified in Eq.~\eqref{eq:energydensity}, used in both numerical examples.}
	\renewcommand*{\arraystretch}{1.15}
	\label{table}
	\begin{tabular}{c|ccc}
		\multirow{2}{*}{Parameter} & $ c_1 $ & $ c_2 $ & $ K $ \\
		& [MPa] & [MPa] & [MPa]\\\hline
		Value & $0.55$ & $0.3$ & $55$ 
	\end{tabular}
\end{table}
%
%
\subsection{Example~1: Local Versus Global Buckling}
The first example analyses a finite column of width~$W$ and height~$H$ with a microstructure consisting of a square stacking of unit cells with edge size~$\ell = 9.97$~mm and circular holes of diameter~$d = 8.67$~mm, cf. Fig.~\ref{fig:squarestacking}. The bottom and top edges of the specimen domain are displaced by~$\pm u/2 \, \vec{e}_2$ to induce~$10\%$ overall compressive strain, defined as~$u/H$. Depending on the slenderness ratio~$H/W$, a competition between microstructural buckling (pattern transformation) and macrostructural buckling of the structure is expected. A similar example has been investigated numerically as well as experimentally by~\cite{Coulais:2015}.

For the DNS solutions, the entire domain is discretized using isoparametric quadratic triangular elements of typical size~$h_\mathrm{m} = \ell/10$ with three Gauss integration points, as shown in Fig.~\ref{fig:columnMacroTriangulation}. For this case, a single microstructural realization adequately represents the ensemble averaged DNS solution. This is shown in Fig.~\ref{fig:columnShifts}, where nominal stress--strain diagrams corresponding to~$100$ microstructural translations (all possible combinations of~$10$ translation steps in horizontal and~$10$ in vertical direction, covering together one period of the microstructure~$\ell \times \ell$) are shown for a~$6\ell \times 30\ell$ specimen. The overall response is initially linear until the first bifurcation point is reached, upon which a local patterning emerges and the specimen's stiffness drops close to zero. Further increasing the compressive strain leads to the second bifurcation, corresponding to global buckling of the specimen, upon which the overall stiffness becomes negative. The maximum and minimum envelopes of all realizations deviate less than~$6\%$ from the corresponding mean, suggesting that the reference microstructure is acceptable for the representation of the effective response.

Only one local patterning mode emerges for the adopted microstructural morphology, i.e.~$n=1$, see Figs.~\ref{fig:DNSlocalvsglobal}--\ref{fig:comphomog} and~\cite{Bertoldi:2008}, approximated analytically as~\cite[see][Eq.~(7)]{Rokos:2019}
\begin{equation}
\begin{aligned}
\vec{\varphi}_1(\vec{X}) 
&= 
\frac{1}{C_1}\left[-\sin\frac{\pi}{\ell}(X_1+X_2) - \sin\frac{\pi}{\ell}(-X_1+X_2) \right]\vec{e}_1 \\
&+
\frac{1}{C_1}\left[\sin\frac{\pi}{\ell}(X_1+X_2) - \sin\frac{\pi}{\ell}(-X_1+X_2) \right]\vec{e}_2,
\end{aligned}
\label{eq:modeSquare}
\end{equation}
were~$C_1$ is a normalization constant ensuring that~$\frac{1}{|Q|}\int_Q\|\vec{\varphi}_1(\vec{X})\|\,\mathrm{d}\vec{X} = 1$, and where~$Q$ is the~$2\ell \times 2\ell$ periodic cell of Fig.~\ref{fig:RVE}. For the RVE discretization, the same type and density of elements is used as for the DNS system (Fig.~\ref{fig:columnMacroTriangulation}). To provide sufficient kinematic freedom to the macroscopic micromorphic system, a mesh convergence study is performed. A uniform macroscopic mesh of quadratic isoparametric triangular elements with three Gauss points is considered with characteristic element sizes~$ h_\mathrm{M} \in \{ 1, \dots, 30 \}\ell $. The same mesh is considered for both macroscopic fields~$\vec{v}_0$ and~$v_1$. An example of a particular mesh of an element size~$h_\mathrm{M} = 2\ell$ for a~$4\ell \times 8\ell$ specimen (for which local buckling is expected to occur) is shown in Fig.~\ref{fig:columnMicroTriangulation}. The obtained results in terms of nominal stress--strain diagrams are plotted in Fig.~\ref{fig:columnMeshConvergence}. For element sizes~$h_\mathrm{M} \leq 4\ell$, the behaviour is similar to the DNS result of Fig.~\ref{fig:columnShifts}; moreover, the results of element sizes~$h_\mathrm{M} = 2\ell$ and~$h_\mathrm{M} = 1\ell$ are indistinguishable. The element size~$h_\mathrm{M} = 2\ell$ is thus adopted in what follows, although from the deformed shape of Fig.~\ref{fig:localvsglobald} it may be clear that a locally refined mesh might be useful. For more details on appropriate choice of element types and associated integration rules see~\cite{Rokos:2020}.
\begin{figure}[htbp]
	\centering
	\subfloat[specimen geometry]{
	\begin{tikzpicture}
	\def\W{4} 
	\def\H{8} 
	\def\cellsize{0.6} 
	\def\diameter{0.867*\cellsize} 
	\coordinate (A) at (0,0);
	\coordinate (B) at (\cellsize*\W,0);
	\coordinate (C) at (\cellsize*\W,\cellsize*\H);
	\coordinate (D) at (0,\cellsize*\H);
	\filldraw[draw = white, fill = blue!5] (A) -- (B) -- (C) -- (D) -- cycle;
	\draw[thin] (D) -- (C);
	\draw[thin] (A) -- (B);
	\draw[thin] (A) -- (D);
	\draw[thin] (B) -- (C);
	\foreach \i in {1,...,\W}{%
		\foreach \j in {1,...,\H}{%
			\filldraw[fill=white,thin] ({\cellsize*(\i-0.5)},{\cellsize*(\j-0.5)}) circle ({\diameter/2});
	}}
	\def\hcons{0.2}
	\foreach \i in {0,0.5,...,\W}{%
	\draw[thin] ({\cellsize*(0+\i)},\cellsize*\H) -- ({\cellsize*(0.15+\i)},{\cellsize*(\H+\hcons)}) -- ({\cellsize*(-0.15+\i)},{\cellsize*(\H+\hcons)}) -- cycle;}
	\draw[thin] (-0.4*\cellsize,{\cellsize*(\H+\hcons)}) -- ({\cellsize*(\W+0.3)},{\cellsize*(\H+\hcons)});
	\foreach \i in {-0.7,-0.45,...,\W}{%
	\draw[thin] ({\cellsize*(\i+0.6)},{\cellsize*(\H+2*\hcons)}) -- ({\cellsize*(\i+0.4)},{\cellsize*(\H+2*\hcons-0.2)});}
	\foreach \i in {0,0.5,...,\W}{%
		\draw[thin] ({\cellsize*(0+\i)},0) -- ({\cellsize*(0.15+\i)},-\cellsize*\hcons) -- ({\cellsize*(-0.15+\i)},-\cellsize*\hcons) -- cycle;}
	\draw[thin] (-0.4*\cellsize,-\cellsize*\hcons) -- ({\cellsize*(\W+0.3)},-\cellsize*\hcons);
	\foreach \i in {-0.7,-0.45,...,\W}{%
		\draw[thin] ({\cellsize*(\i+0.4)},-\cellsize*\hcons) -- ({\cellsize*(\i+0.6)},{\cellsize*(-\hcons-0.2)});}
	\draw[thin,latex'-latex'] (-0.65,0) -- (-0.65,\cellsize*\H) node[midway,left] {\footnotesize $H$};
	\draw[thin] (-0.8,0) -- (-0.1,0);
	\draw[thin] (-0.8,\cellsize*\H) -- (-0.1,\cellsize*\H);
	\draw[thin,latex'-latex'] (0,\cellsize*\H+0.45) -- (\cellsize*\W,\cellsize*\H+0.45) node[midway,above] {\footnotesize $W$};
	\draw[thin] (0,{\cellsize*(\H+0.5)}) -- (0,{\cellsize*(\H+1)});
	\draw[thin] (\cellsize*\W,{\cellsize*(\H+0.5)}) -- (\cellsize*\W,{\cellsize*(\H+1)});
	\draw[thin,-latex'] (-0.2,\cellsize*\H) -- (-0.2,{\cellsize*(\H-1)}) node[midway,left] {\footnotesize $\frac{u}{2}$};
	\draw[thin,-latex'] (-0.2,0) -- (-0.2,\cellsize) node[midway,left] {\footnotesize $\frac{u}{2}$};	
	\node[anchor=north east,shift={(-0.05*\cellsize,-0.05*\cellsize)}] (domain) at (C) {\footnotesize $\Omega$};	
	\draw[-latex',line width=0.3mm] (0.5*\W*\cellsize,0.5*\H*\cellsize) -- (0.5*\W*\cellsize,0.5*\H*\cellsize+1) node[anchor = east,shift={(-0.0,-0.1)}] {\footnotesize $\vec{e}_{2}$};
	\draw[-latex',line width=0.3mm] (0.5*\W*\cellsize,0.5*\H*\cellsize) -- (0.5*\W*\cellsize+1,0.5*\H*\cellsize) node[anchor = south,shift={(-0.05,0)}] {\footnotesize $\vec{e}_{1}$};
	\draw[thin,latex'-latex'] ({\cellsize*(\W+0.5)},\cellsize) -- ({\cellsize*(\W+0.5)},2*\cellsize) node[midway,right] {\footnotesize $\ell$};
	\draw[dashed] ({\cellsize*(\W+0.5)},\cellsize) -- ({\cellsize*(\W-1)},\cellsize) -- ({\cellsize*(\W-1)},2*\cellsize) -- ({\cellsize*(\W+0.5)},2*\cellsize);
	\draw[thin,latex'-latex'] ({\cellsize*(\W-0.5)},{\cellsize*(1+0.0665)}) -- ({\cellsize*(\W-0.5)},{\cellsize*(2-0.0665)}) node[midway,right,shift={(-0.12,0)}] {\footnotesize $d$};
	\end{tikzpicture}\label{fig:squarestacking}}
	\hspace{1em}
	\subfloat[micro-mesh]{\includegraphics[scale=1]{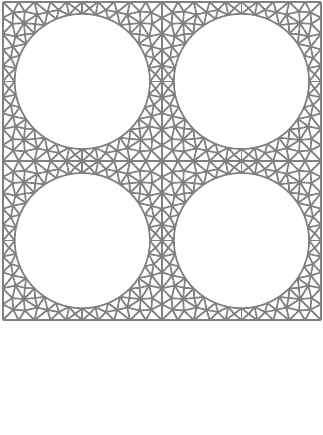}\label{fig:columnMacroTriangulation}}
	\hspace{1em}	
	\subfloat[macro-mesh]{\includegraphics[scale=0.865]{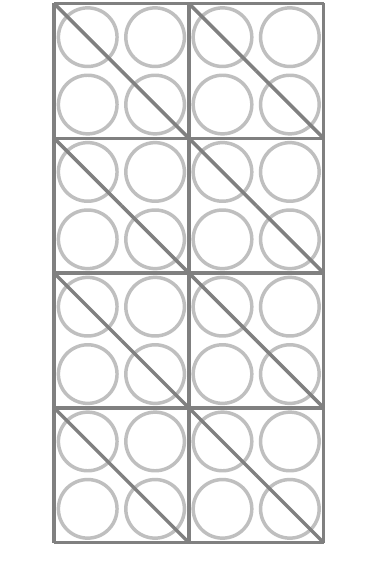}\label{fig:columnMicroTriangulation}}
	\caption{(a)~Specimen geometry for the local versus global buckling example, i.e.~a specimen of a width~$W = 4\ell$ and height~$H = 8\ell$ made of an elastomeric mechanical metamaterial, subjected to an overall vertical compressive strain of~$10\%$. (b)~Typical discretization of the full Direct Numerical Simulation~(DNS) and RVE model, element size~$h_\mathrm{m} = \ell/10$. (c)~Macroscopic discretization of a~$4\ell \times 8\ell$ specimen with element size~$h_\mathrm{M} = 2\ell$. In both cases, isoparametric quadratic triangular elements with three Gauss points are used.}
	\label{fig:columnGeometry}
\end{figure}
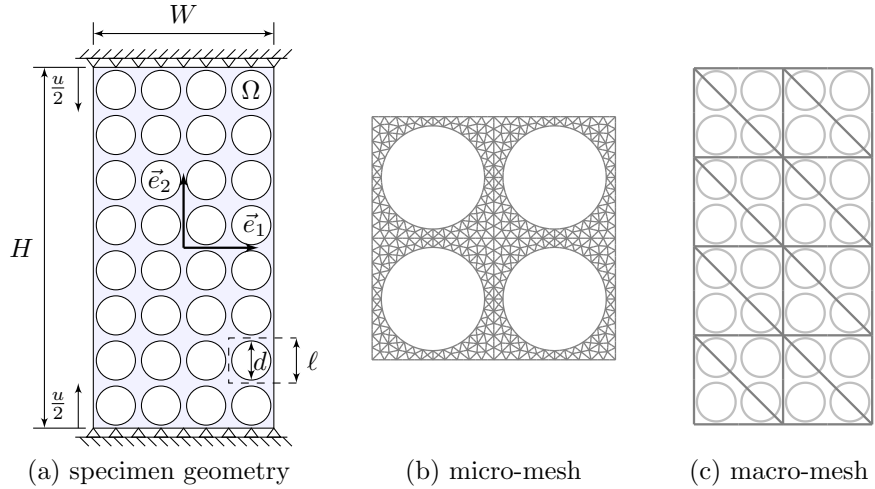
\begin{figure}[htbp]
	\centering
	\subfloat[DNS shifts]{\includegraphics[scale=1]{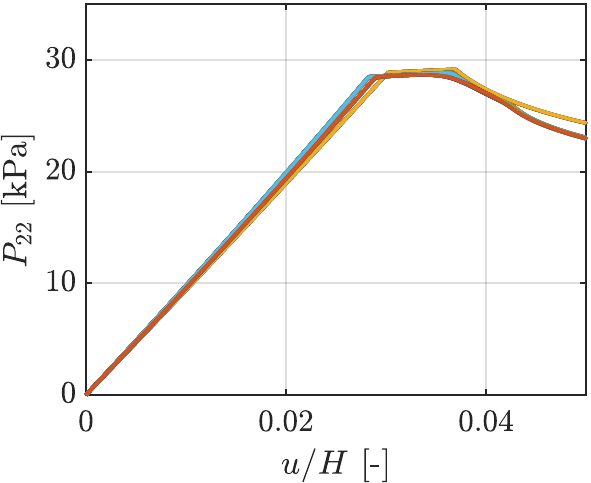}\label{fig:columnShifts}}
	\hspace{2em}	
	\subfloat[MM mesh convergence study]{\includegraphics[scale=1]{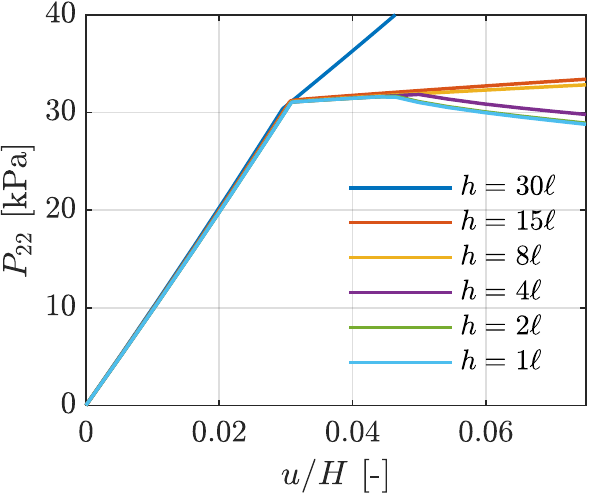}\label{fig:columnMeshConvergence}}
	\caption{(a)~Stress--strain diagrams for $100$ translated DNS solutions ($10$ translation steps in horizontal and~$10$ in vertical direction, covering one period of the microstructure~$\ell \times \ell$), $W = 6\ell$ and~$H = 30\ell$. (b)~Macroscopic mesh convergence study for the micromorphic~(MM) homogenization scheme with a uniform triangulation, cf. Fig.~\ref{fig:columnMicroTriangulation}. Considered element sizes~$ h \in \{ 1, \dots, 30 \} \ell $ for~$ W = 6 \ell $ and~$ H = 30 \ell $ (slenderness ratio~$H/W = 5$).}
	\label{fig:columnConvergence}
\end{figure}

Depending on the slenderness ratio~$H/W$, two basic and mutually interacting deformation mechanisms occur, as shown in Fig.~\ref{fig:localvsglobal} by the deformed configurations for~$ 6\ell \times 14\ell $ and~$ 6\ell \times 34\ell $ specimens. The first mechanism is local patterning (Fig.~\ref{fig:localvsglobala}), emerging for low slenderness ratios upon reaching a critical compressive strain of approximately~$3\%$. The cells fold in a typical pattern of alternating ellipsoidal holes and an auxetic effect is observed along with boundary layers where the local buckling is restricted. The second mechanism, occurring for higher slenderness ratios, is global buckling (Fig.~\ref{fig:localvsglobalc}), which is triggered upon reaching the critical Euler buckling stress. The corresponding buckling strain can be estimated as
\begin{equation}
\varepsilon_\mathrm{cr}(H/W) = \frac{\pi^2}{3 (H/W)^2}.
\label{eq:eulerBuckling}
\end{equation}

In Figs.~\ref{fig:localvsglobalb} and~\ref{fig:localvsglobald}, the micromorphic field normalized by its maximum value considered over space and a parametrization pseudo-time~$t$, i.e.~$\widehat{v}_1 = v_1 / \| v_1(t,\vec{X}) \|_\infty$, is shown in colour. Comparing Fig.~\ref{fig:localvsglobala} with~\ref{fig:localvsglobalb}, and Fig.~\ref{fig:localvsglobalc} with~\ref{fig:localvsglobald}, we conclude that the micromorphic homogenization scheme is capable of accurately reconstructing the overall kinematic response, correctly capturing the auxetic effect for lower slenderness ratios, and accurately indicating regions of localized patterning reflected by the magnitude of the micromorphic field~$\widehat{v}_1$ for larger slenderness ratios. The pattering regions localize in the compressive parts of the bent domain, situated close to the specimen's centre and near the supports at both ends. The overall deformed shape for the large slenderness ratio, i.e.~the~$\vec{v}_0$ field, is captured with good accuracy as well.

Nominal stress--strain diagrams for~$W = 6\ell$ and slenderness ratios~$H/W \in [ 1, 30 ]$ are shown in Fig.~\ref{fig:stressstrain}. Here, the two mechanisms and their mutual interactions are visible more clearly. For the DNS results (Fig.~\ref{fig:stressstraina}) and the applied strain range~$u/H \in [0,0.1]$ the slenderness ratios up to~$H/W = 2.33$ buckle only locally, ratios~$H/W \in [2.67,6.67]$ buckle first locally and then globally, the ratio~$H/W = 7$ buckles locally and globally at the same time, whereas ratios~$H/W \in [7.33,30]$ buckle first globally and then locally. Bilinear stress--strain responses typically emerge, which exhibit softening in later stages due to the presence of secondary (local or global) buckling, see also Fig.~\ref{fig:columnShifts} and the discussion therein. A close-up on the local versus global buckling intersection is shown in Fig.~\ref{fig:stressstrainzooma}, where mild snap-backs for slenderness ratios~$H/W \in [ 6.33, 8.33 ]$ can be observed. The micromorphic computational homogenization is capable of reconstructing the stability behaviour (Fig.~\ref{fig:stressstrainb}) with an accuracy that decreases with decreasing scale ratio (i.e.~larger errors are observed for smaller scale ratios). In particular, for~$H/W = 1$ we see a large discrepancy in the post-bifurcation nominal stiffness (i.e.~in the slope of the~$P_{22}$ versus~$u/H$ curve), which corresponds to~$25\%$ of error relative to the initial pre-bifurcation stiffness (which is practically constant for all considered~$H/W$ ratios). With increasing slenderness ratio, however, the error drops rapidly down to~$0.1\%$. For better clarity, the bifurcation curves corresponding to the DNS and micromorphic results are compared in Fig.~\ref{fig:stressstrainzoomb}. Here the shapes as well as slopes of the DNS bifurcation curves (shown in blue) are captured accurately by the micromorphic scheme (shown in red), although micromorphic homogenization systematically overestimates the DNS results. The maximum relative error in terms of the critical buckling stress of the first instability is of the order of~$12\%$, but not lower than~$7\%$ even for large scale ratios.
\begin{figure}[htbp]
	\centering
	\subfloat[DNS, $6\ell \times 14\ell$]{\includegraphics[scale=1]{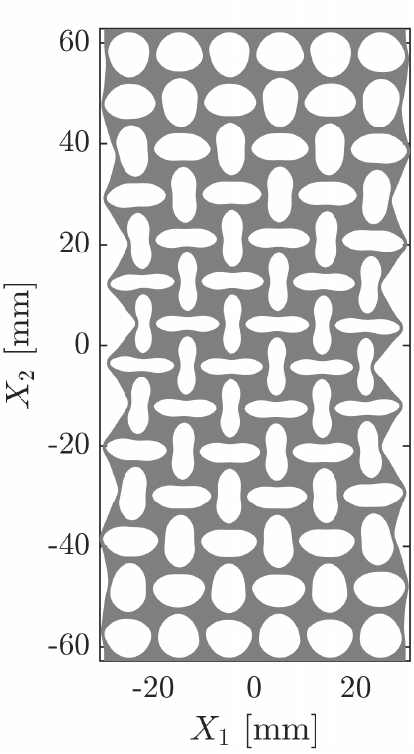}\label{fig:localvsglobala}}	
	\hspace{0.4em}
	\subfloat[MM, $6\ell \times 14\ell$]{\includegraphics[scale=1]{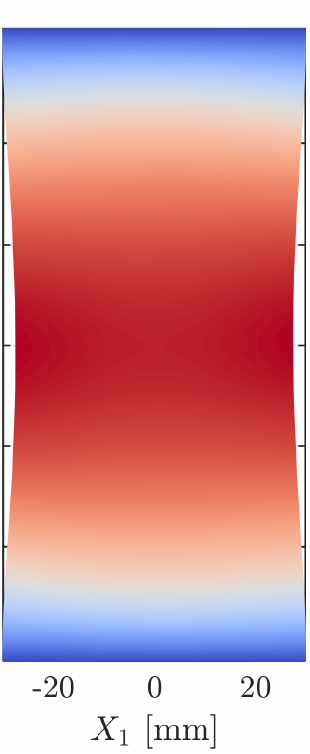}\label{fig:localvsglobalb}}
	\hspace{2em}
	\subfloat[DNS, $6\ell \times 34\ell$]{\includegraphics[scale=1]{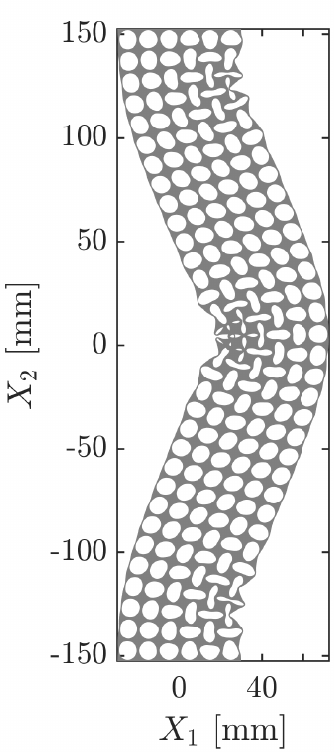}\label{fig:localvsglobalc}}
	\hspace{0.5em}	
	\subfloat[MM, $6\ell \times 34\ell$]{\includegraphics[scale=1]{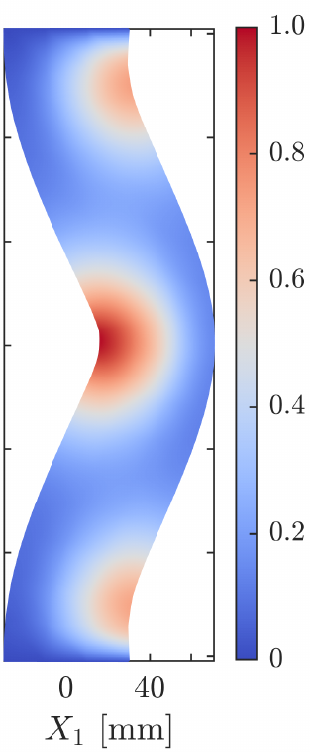}\label{fig:localvsglobald}}	
	\caption{Comparison of deformed shapes obtained from Direct Numerical Simulation~(DNS) and the micromorphic~(MM) computational homogenization scheme. (a)~DNS and~(b) MM results for a~$6\ell \times 14\ell$ specimen (slenderness ratio~$H/W = 2.333$, $\| v_1(t,\vec{X}) \|_\infty = 51.5$). (c)~DNS and~(d) MM results for a~$6\ell \times 34\ell$ specimen ($H/W = 5.667$, $\| v_1(t,\vec{X}) \|_\infty = 69.5$). Local buckling emerges in~(a) and~(b), whereas global buckling is found in~(c) and~(d). The colour in the MM plots indicates the magnitude of the~$v_1$ field normalized by its extreme value in time and space, i.e.~$\widehat{v}_1$. The results are shown for an overall applied strain~$u/H = 0.1$.}
	\label{fig:localvsglobal}
\end{figure}
\begin{figure}[htbp]
	\centering
	\subfloat[DNS, stress--strain curves]{\includegraphics[scale=1]{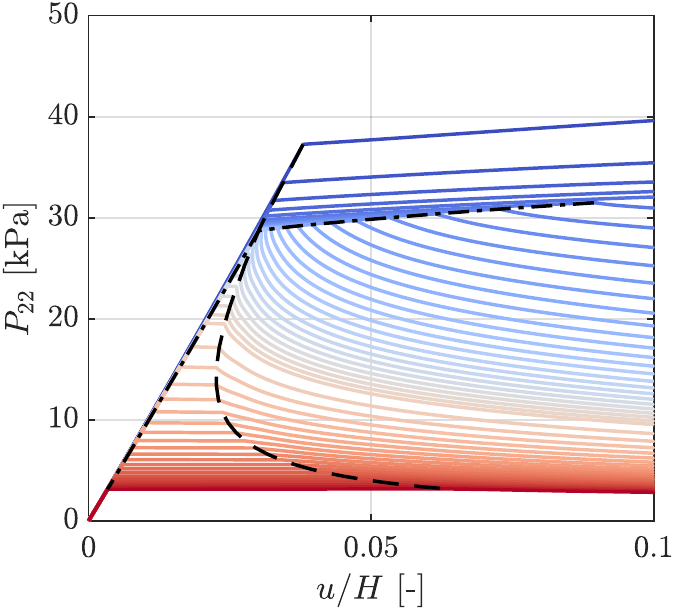}\label{fig:stressstraina}}
	\hspace{0.5em}
	\subfloat[MM, stress--strain curves]{\includegraphics[scale=1]{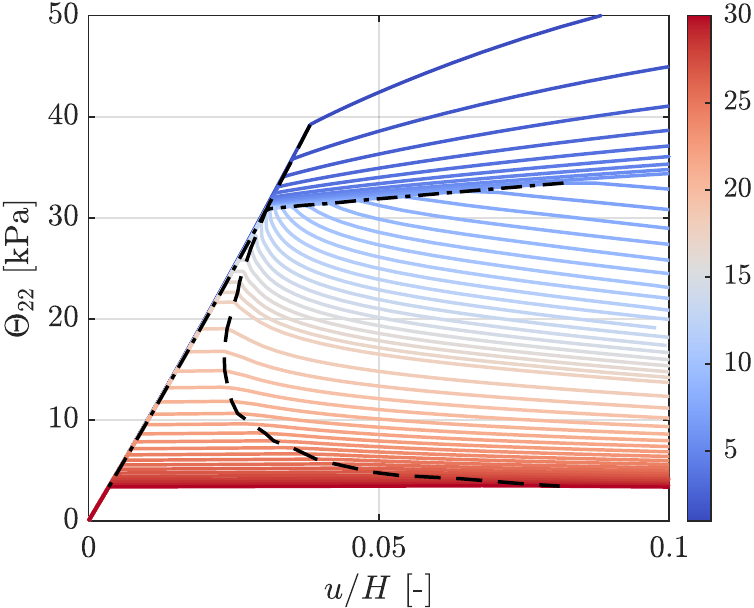}\label{fig:stressstrainb}}
	\caption{The nominal stresses~$P_{22}$ and~$\Theta_{22}$ as a function of the applied nominal strain~$u/H$ for columns of various slenderness ratios~$H/W \in [ 1, 30 ]$, $ W = 6\ell $, shown in colour, obtained via~(a) Direct Numerical Simulation~(DNS), and~(b) micromorphic~(MM) computational homogenization. Instability points corresponding to global buckling are connected by the black dash-dot lines (\sampleline{dash pattern=on 0.7em off 0.2em on 0.05em off 0.2em}), whereas local buckling points are connected by the black dashed lines (\sampleline{dash pattern=on 0.5em off 0.25em}); cf.~also Fig.~\ref{fig:stressstrainzoom}.}
	\label{fig:stressstrain}
\end{figure}
\begin{figure}[htbp]
	\centering
	\subfloat[DNS stress--strain curves, close-up]{\includegraphics[scale=1]{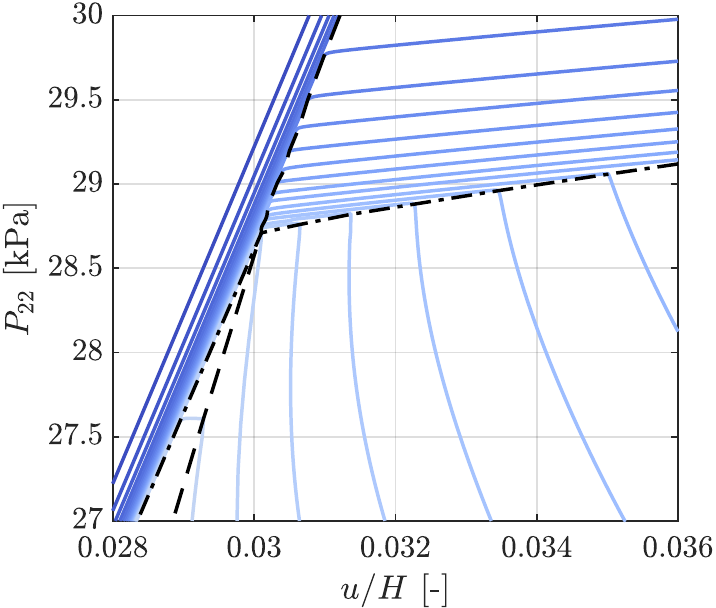}\label{fig:stressstrainzooma}}
	\hspace{1em}
	\subfloat[bifurcation curves]{\includegraphics[scale=1]{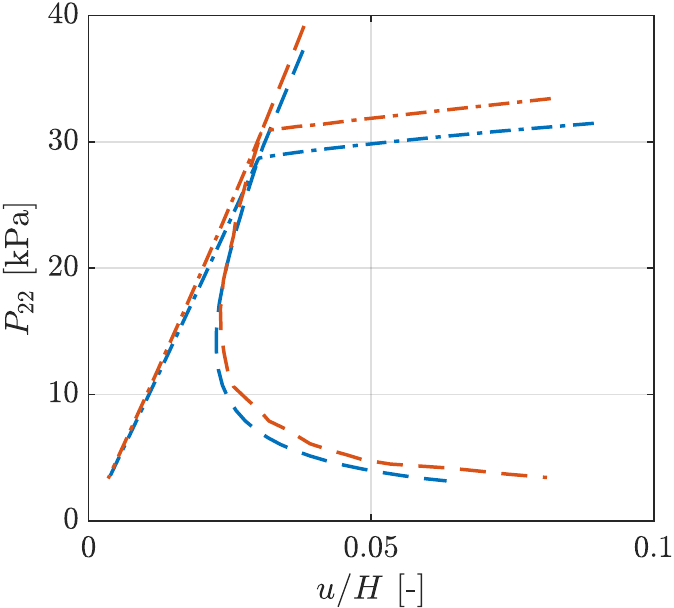}\label{fig:stressstrainzoomb}}
	\caption{(a)~A close-up on stress--strain curves corresponding to the DNS solutions of Fig.~\ref{fig:stressstraina} in the vicinity of the local versus global buckling intersection. (b)~Bifurcation curves for direct numerical solutions (blue, cf. Fig.~\ref{fig:stressstraina}) and micromorphic homogenization (red, cf. Fig.~\ref{fig:stressstrainb}). Instability points corresponding to global buckling are connected by the dash-dot lines (\sampleline{dash pattern=on 0.7em off 0.2em on 0.05em off 0.2em}), whereas local buckling points are connected by the dashed lines (\sampleline{dash pattern=on 0.5em off 0.25em}).}
	\label{fig:stressstrainzoom}
\end{figure}

\begin{figure}[htbp]
	\centering
	\includegraphics[scale=1]{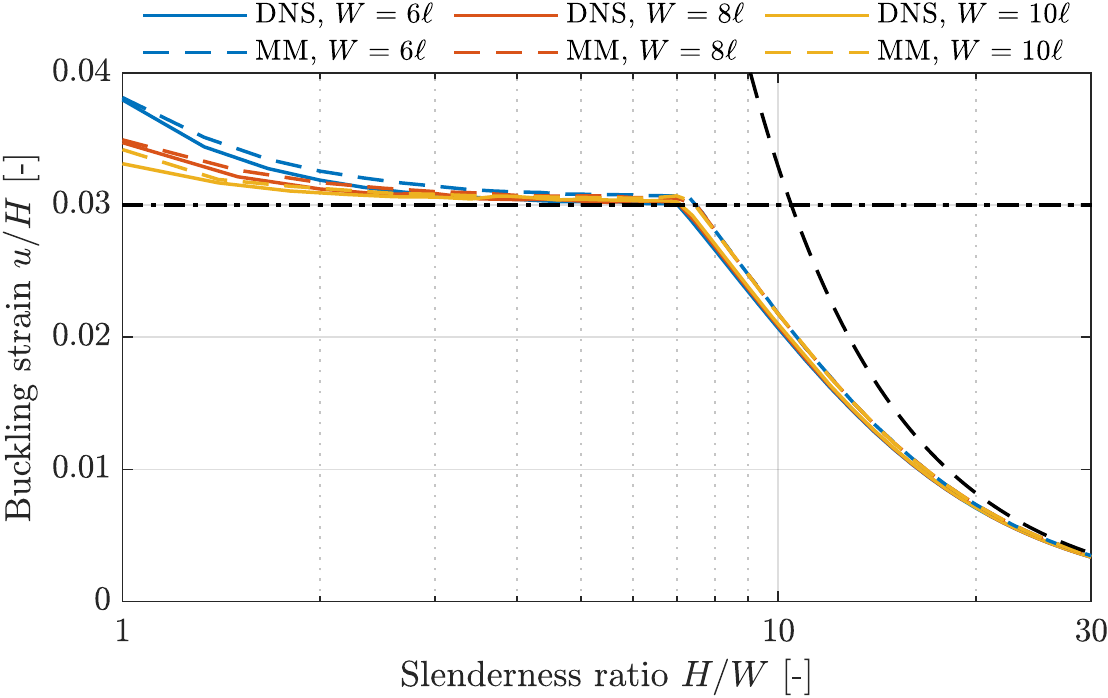}
	\caption{Buckling strain $u/H$ as a function of the slenderness ratio~$H/W \in [1,30]$, $W \in \{6, 8, 10\}\ell$, obtained from the DNS and micromorphic~(MM) computational homogenization algorithms, and from the theoretical estimates for local (\sampleline{dash pattern=on 0.7em off 0.2em on 0.05em off 0.2em}) and global (\sampleline{dash pattern=on 0.5em off 0.25em}) buckling.}
	\label{fig:strainslenderness}
\end{figure}

The buckling strains expressed as a function of the slenderness ratio~$H/W$ and corresponding to the first bifurcation points of Fig.~\ref{fig:stressstrain} are shown in Fig.~\ref{fig:strainslenderness} for several specimen widths~$W \in \{6, 8, 10\}\ell$. \cite{Bertoldi:2008} reported that the buckling strain of a single RVE corresponds to approximately~$3\%$ (shown as the black dash-dot line in Fig.~\ref{fig:strainslenderness}), and hence this value is expected to be the theoretical local buckling strain. Both the DNS as well as micromorphic results attain this limit for the range of slenderness ratios~$H/W \in [1,7]$, although for very small slenderness ratios a mild increase in the critical strain is observed. This effect is explained by the growing influence of the stiff boundary layers constraining the evolution of the microstructural patterns. Note that short columns may still buckle globally, upon further increase of the external load. For a slenderness ratio of approximately~$H/W = 7$, local and global buckling occur simultaneously, whereas higher slenderness ratios converge asymptotically towards the theoretical bound of the global Euler buckling strain given by Eq.~\eqref{eq:eulerBuckling} (shown as the black dashed line). Both the DNS and the micromorphic scheme approach this limit from below, although the micromorphic results overestimate systematically the critical buckling strain obtained by the DNS. With increasing width of the specimen~$W$, the size effects present for small slenderness ratios slowly decrease. For short columns the local buckling strain converges towards the theoretical bound~$3\%$, whereas the global buckling strain shows little change. The overall relative error of the micromorphic scheme in terms of the buckling strain does not exceed~$3\%$ for local and~$6\%$ for global buckling. With increasing scale ratio, this error drops down to~$0.5\%$ for local and~$3\%$ for global buckling. Note that the theoretical global Euler buckling strain given by Eq.~\eqref{eq:eulerBuckling} significantly overestimates the DNS results due to a substantial amount of shear and changes triggered in the microstructure upon buckling (cf. Fig.~\ref{fig:localvsglobalc}), which become important especially for intermediate slenderness ratios~$H/W \in (7,20]$.
%
%
\subsection{Example~2: Multiple Local Modes}
\label{sect:hexagonal}
\begin{figure}
	\centering
	\subfloat[hexagonal stacking]{
	\begin{tikzpicture}
		\def\W{6};
		\def\H{6};
		\def\scale{0.88}
		\def\factor{0.9235};
		\clip (\scale*0.5,\scale*0.5) rectangle(\scale*5.5,\scale*5);
		\coordinate (A) at (0,0);
		\coordinate (B) at (\scale*\W,0);
		\coordinate (C) at (\scale*\W,\scale*\H);
		\coordinate (D) at (0,\scale*\H);
		\filldraw[draw = white, fill = blue!5] (A) -- (B) -- (C) -- (D) -- cycle;
		\draw[thin] (D) -- (C);
		\draw[thin] (A) -- (B);
		\draw[thin] (A) -- (D);
		\draw[thin] (B) -- (C);
		\foreach \i in {1,...,\W}{%
			\foreach \j in {1,3,...,\H}{%
				\filldraw[fill=white,thin] ({\scale*(\i-0.5)},{\scale*(\j*\factor-0.5*\factor)}) circle (\scale*0.9235/2);}}
		\foreach \i in {1.5,2.5,...,\W}{%
			\foreach \j in {2,4,...,\H}{%
				\filldraw[fill=white,thin] ({\scale*(\i-0.5)},{\scale*(\j*\factor-0.5*\factor)}) circle (\scale*0.9235/2);}}
	\end{tikzpicture}\label{fig:hexaa}}
	\hspace{1em}
	\subfloat[hexagonal RVE]{\def\svgwidth{3.75cm}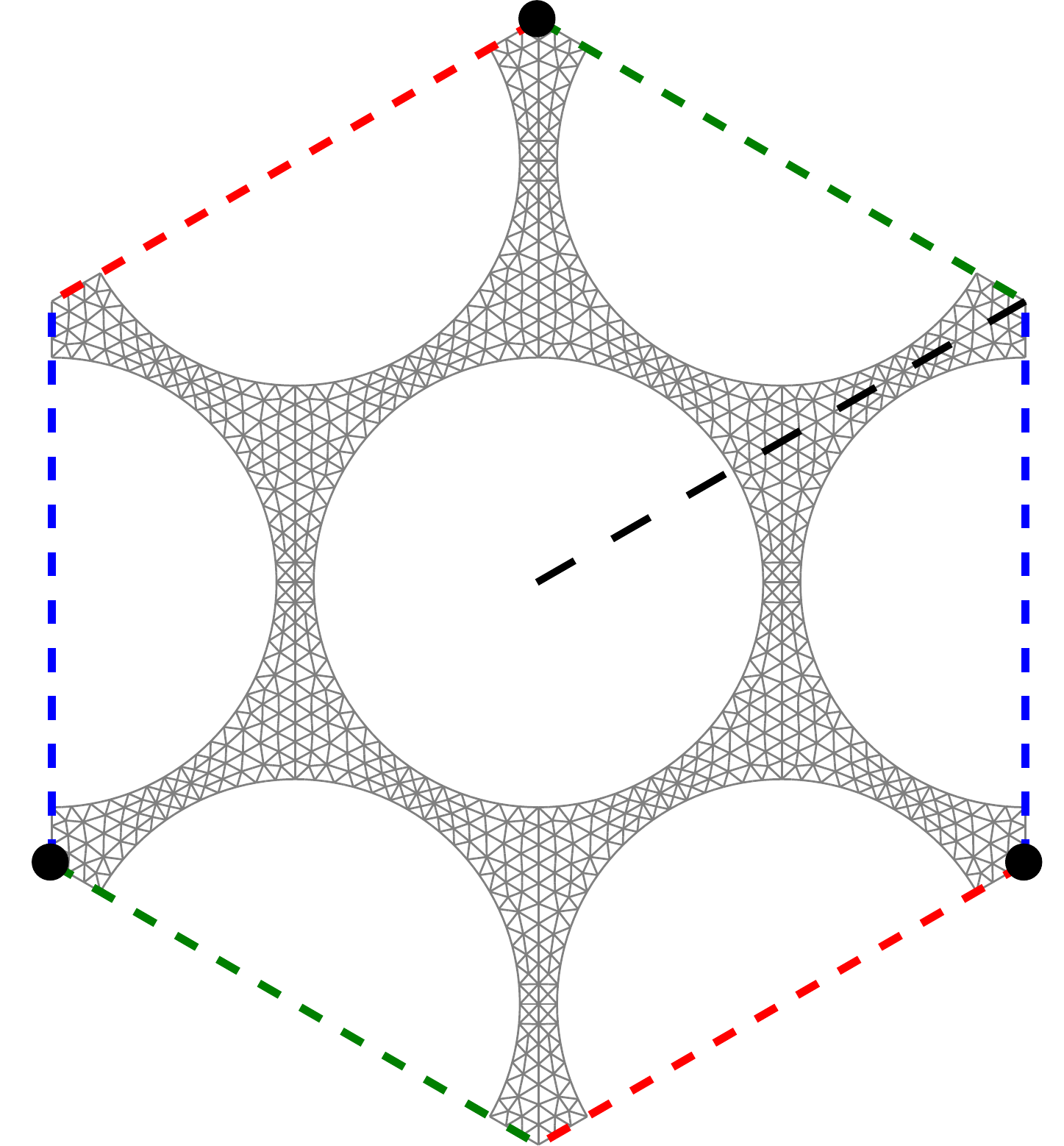\label{fig:hexab}}
	\hspace{1em}
	\subfloat[macroscopic mesh]{\includegraphics[scale=0.975]{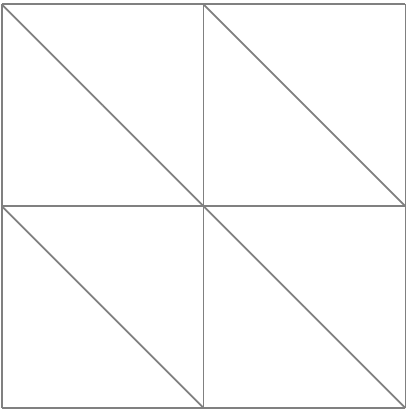}\label{fig:hexac}}	
	\caption{(a)~A microstructure with a hexagonal stacking of holes in the reference configuration. (b)~A single hexagonal RVE with an average mesh size~$h_\mathrm{m} = \ell/10$. The boundaries indicated by the same colour are coupled via the periodicity constraint~\eqref{eq:periodicity}. (c)~A macroscopic periodic mesh for the micromorphic computational homogenization scheme, $h_\mathrm{M} = 3.6\ell$.}
	\label{fig:hexa}
\end{figure}

\begin{figure}[h]
	\centering
	\subfloat[pattern~I]{\includegraphics[height=4cm]{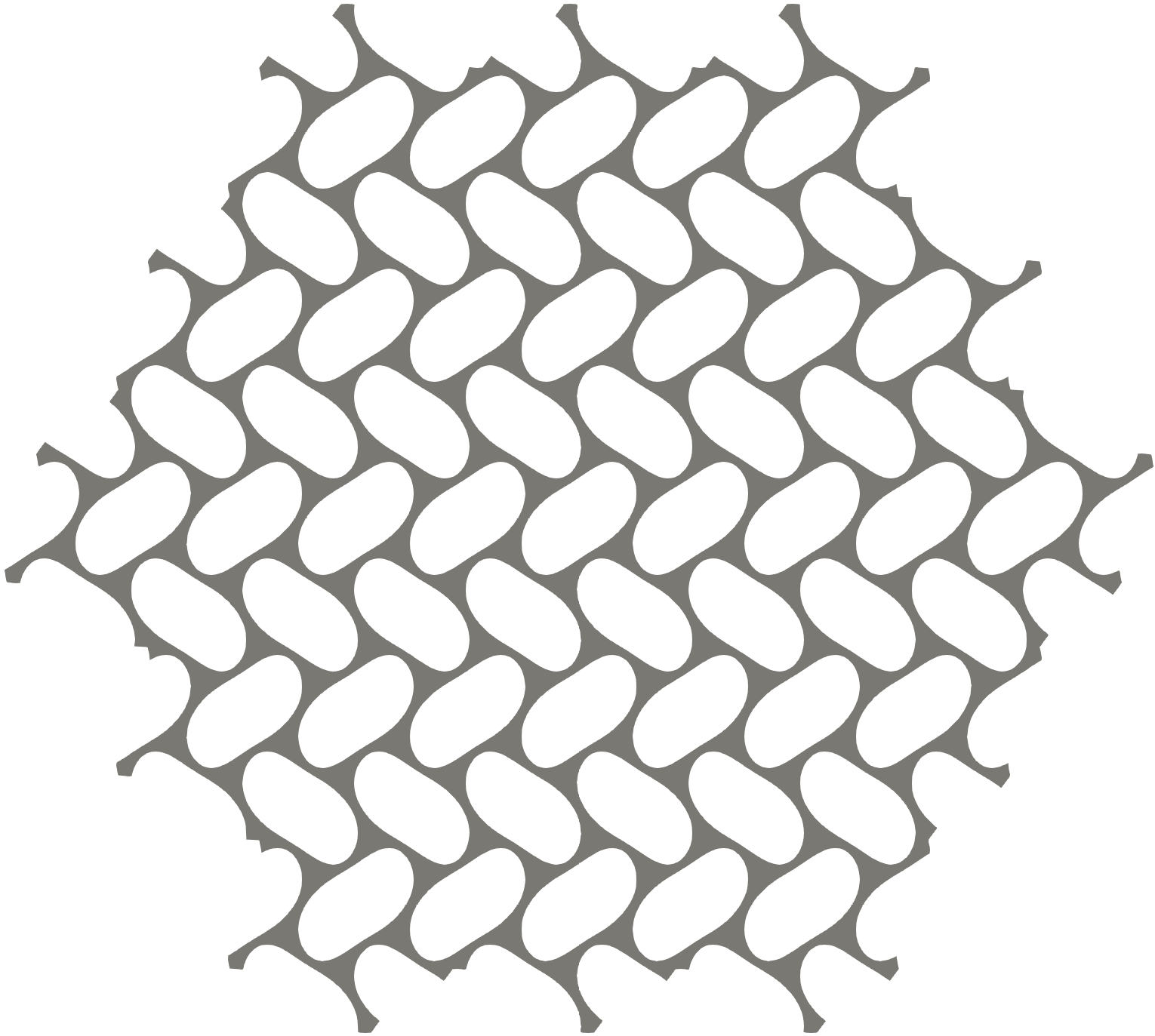}\label{fig:pattern1}}
	\hspace{1em}
	\subfloat[pattern~II]{\includegraphics[height=4cm]{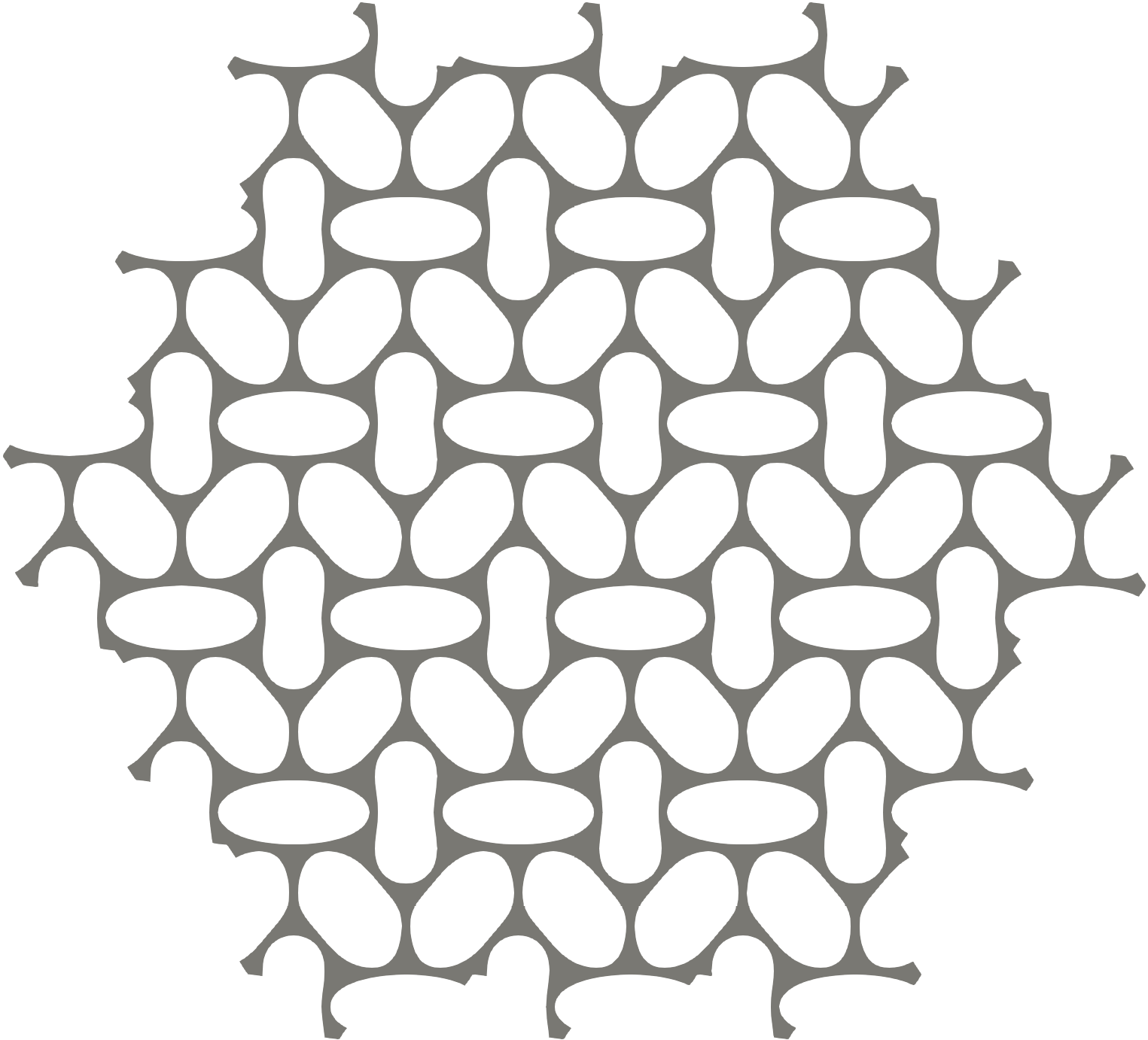}\label{fig:pattern2}}
	\hspace{1em}
	\subfloat[pattern~III]{\includegraphics[height=4cm]{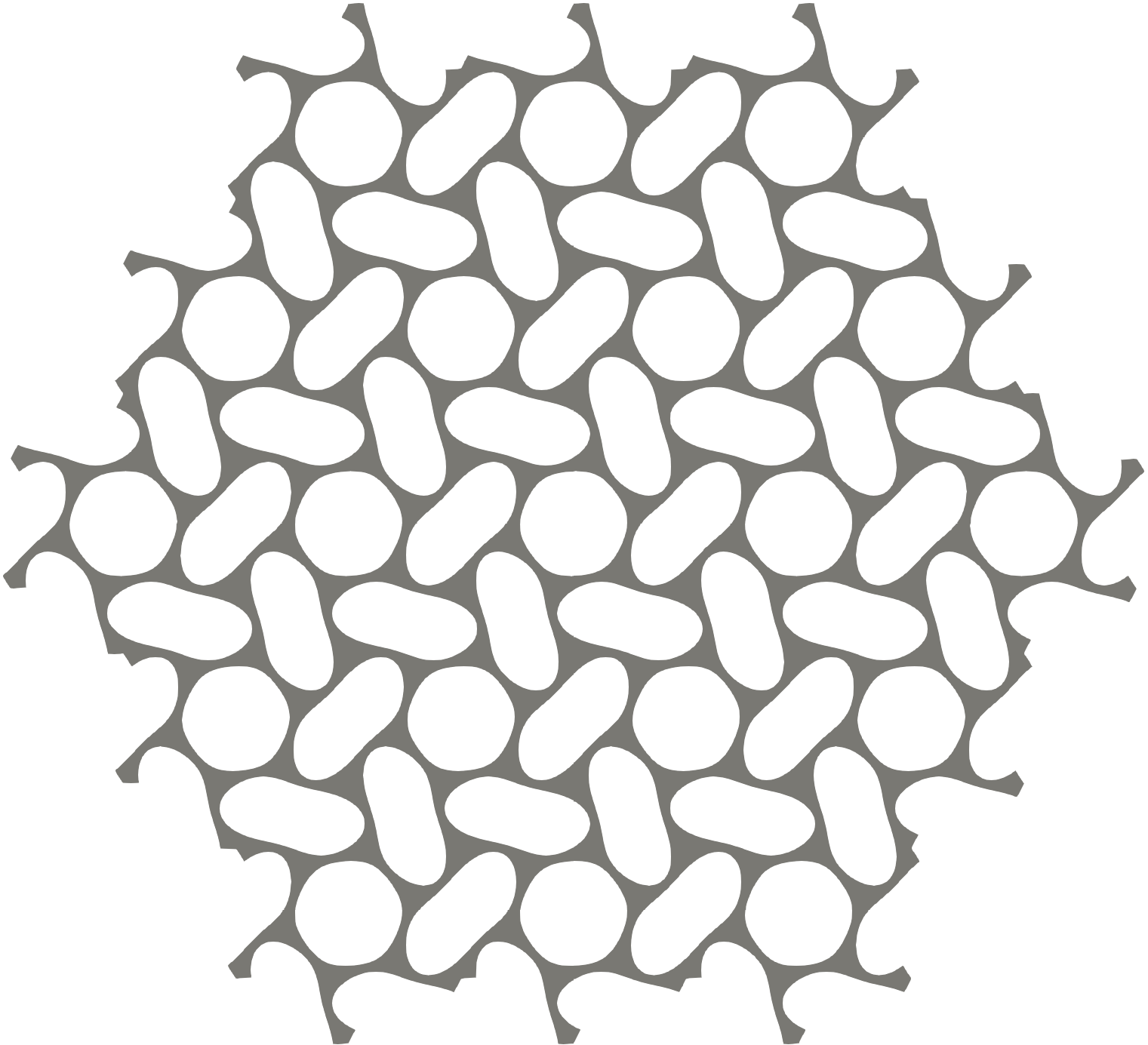}\label{fig:pattern3}} \\
	\subfloat[mode~I]{\includegraphics[height=4cm]{mode_I-crop.png}\label{fig:mode1}}
	\hspace{1em}
	\subfloat[mode~II]{\includegraphics[height=4cm]{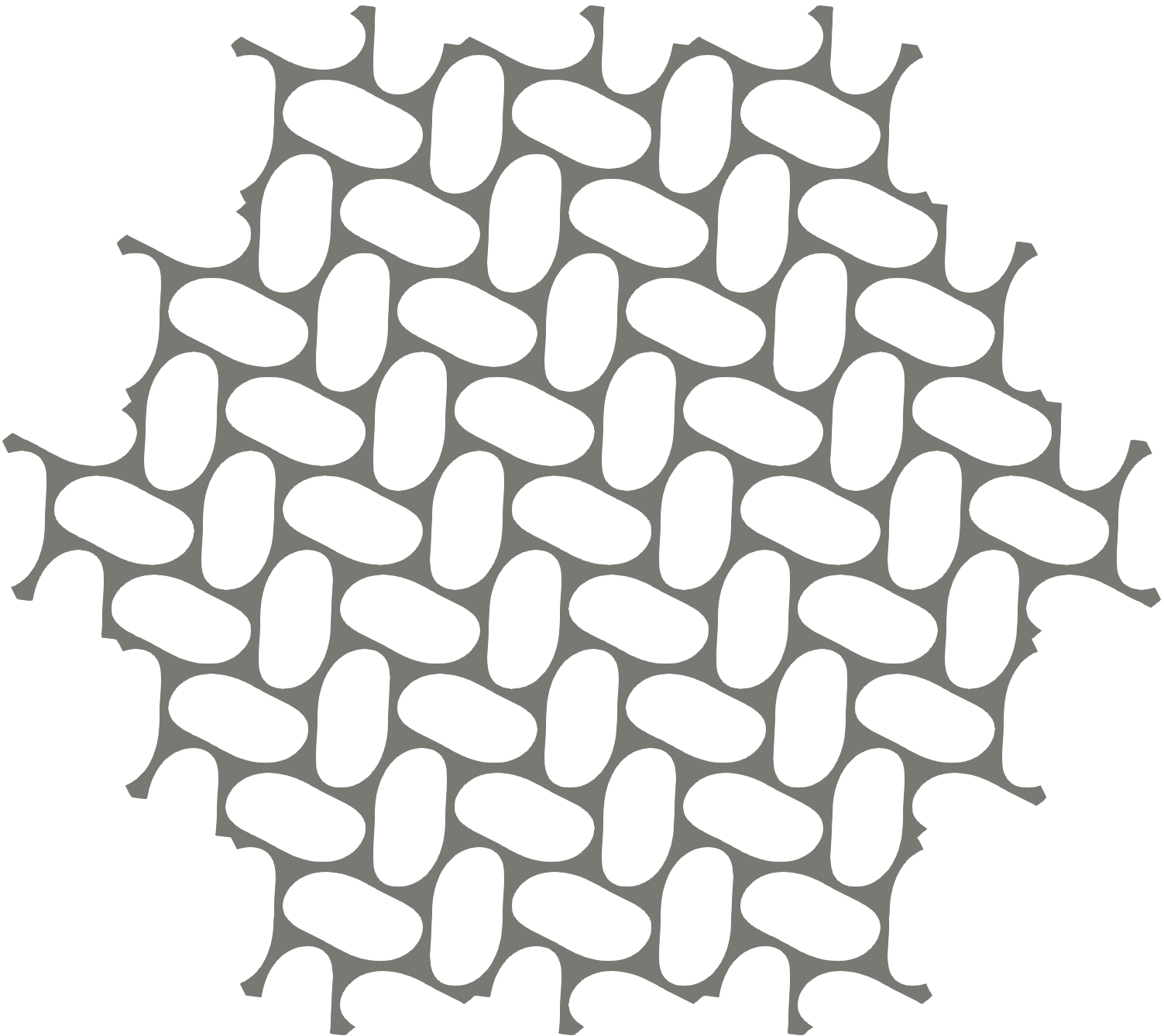}\label{fig:mode2}}
	\hspace{1em}
	\subfloat[mode~III]{\includegraphics[height=4cm]{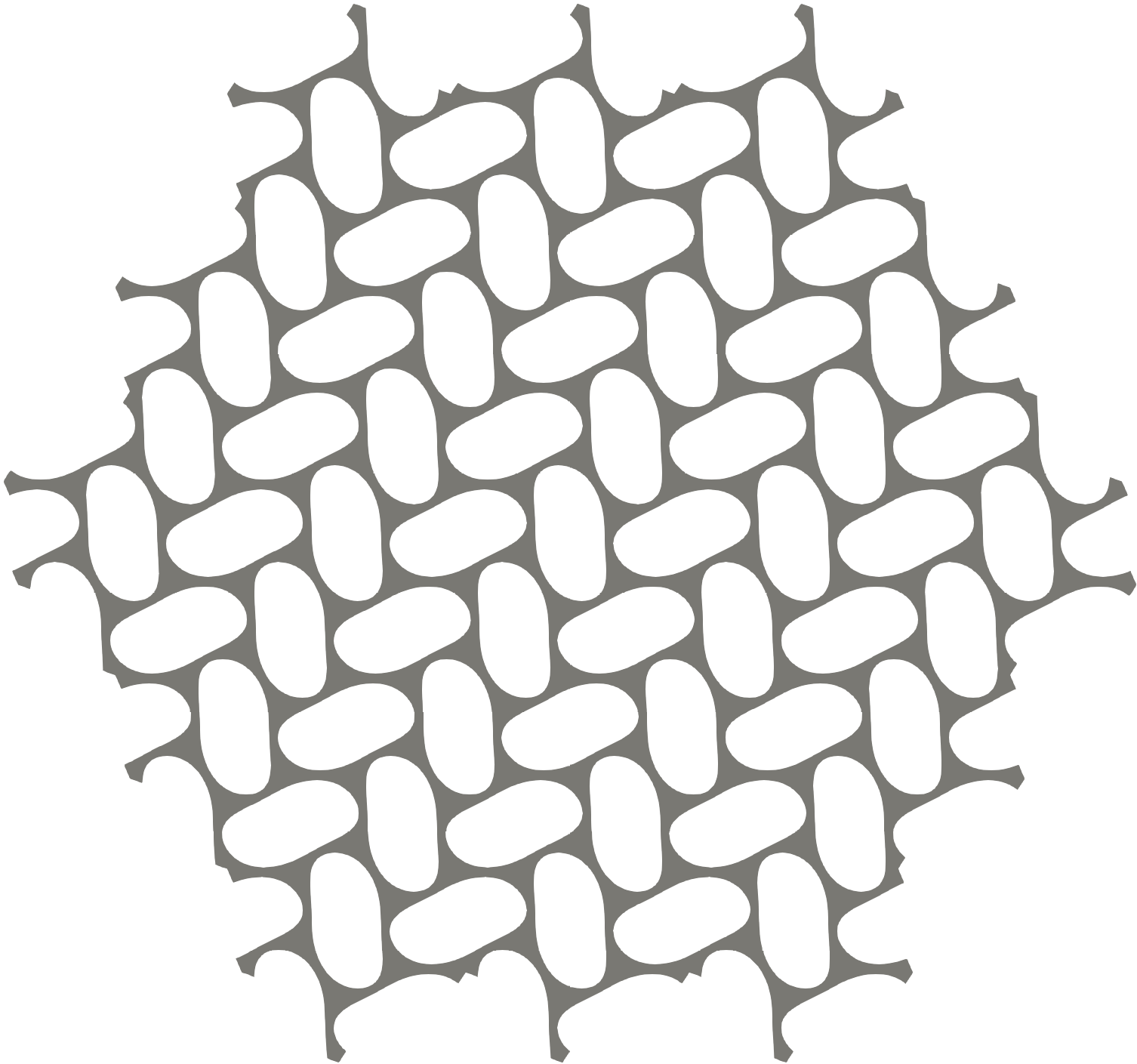}\label{fig:mode3}}	
	\caption{Three pattern transformations~$\vec{\pi}_i$ and orthogonal modes~$\vec{\varphi}_i$. (a) Pattern I, $\vec{\pi}_1$, the uniaxial or shear pattern, corresponds to $\gamma > 1$. (b) Pattern II, $\vec{\pi}_2$, the biaxial or butterfly pattern, occurs for $\gamma < 1$. (c)~Pattern III, $\vec{\pi}_3$, the equi-biaxial pattern or flower-like pattern, emerges for $\gamma = 1$. (d)~Mode~I, $\vec{\varphi}_1$, developing perpendicular to~$\theta = 90^\circ$, (e) mode~II, $\vec{\varphi}_2$, developing perpendicular to~$\theta = 30^\circ$, and~(f) mode~III, $\vec{\varphi}_3$, developing perpendicular to~$\theta = - 30^\circ$.}
	\label{fig:patterns}
\end{figure}

The second example considers an infinite microstructure composed of a hexagonal stacking of holes, shown in Fig.~\ref{fig:hexaa}. The periodic cell, considered later as RVE for the micromorphic scheme, thus comprises two holes in each of the three directions along the hole centres, see Fig.~\ref{fig:hexab}. As reported by \cite{Ohno:2002} for the case of hexagonal honeycombs, three different patterns can emerge under biaxial compression, depending on the biaxiality ratio
\begin{equation}
\gamma = \frac{|\overline{F}_{22}-1|}{|\overline{F}_{11}-1|} = \frac{|\varepsilon_{22}|}{|\varepsilon_{11}|} \in [0,\infty],
\label{eq:biaxiality}
\end{equation}
where~$\varepsilon_{ii}=\overline{F}_{ii}-1$ are the nominal compressive normal strains in the~$\vec{e}_i$, $i = 1,2$, directions, and~$\overline{F}_{ij}$ are the components of the overall deformation gradient tensor~$\overline{\bs{F}}$, with~$\overline{F}_{12} = \overline{F}_{21} = 0$. The three distinct patterns, depicted in Figs.~\ref{fig:pattern1}--\ref{fig:pattern3}, correspond to the following cases: (i)~\emph{Pattern~I}, uniaxial or shear pattern denoted~$\vec{\pi}_1$, occurs when the compressive load on the vertical cell walls is higher than the load on the other cell walls, i.e.~$\gamma > 1$. The multiplicity of the bifurcation point corresponds to one. The displacement field leads to the formation of horizontal layers of holes sheared alternatingly to the right and to the left, see Fig.~\ref{fig:pattern1}. (ii)~\emph{Pattern II}, also called biaxial or butterfly-like pattern and denoted~$\vec{\pi}_2$, emerges when the inclined cell walls at~$\theta = \pm 30^{\circ}$ are compressed more than the vertical cell walls, i.e.~$\gamma <1$. In this case, the multiplicity of the bifurcation point is two and the pattern exhibits horizontal layers of holes buckled along the horizontal and vertical directions, see Fig.~\ref{fig:pattern2}. (iii) \emph{Pattern III}, also referred to as the equi-biaxial or flower-like pattern~$\vec{\pi}_3$, is observed when all three cell walls are subjected to an equal compressive load, i.e.~$\gamma = 1$. The multiplicity of the bifurcation point equals three, and the displacement field corresponds to a virtually undeformed central hole, surrounded by ellipses, see Fig.~\ref{fig:pattern3}.

Because the individual patterns~$\vec{\pi}_i$ are not mutually orthogonal, it is convenient for further treatment to introduce the so-called modes~$\vec{\varphi}_i$, $i = 1, 2, 3$, \cite[see][]{Ohno2002,Okumura2002,Rokos:2019a}, which satisfy orthogonality. Linear combinations of these modes result in the previously introduced patterns as follows:
\begin{align}
\vec{\pi}_1 &= \vec{\varphi}_1, \label{eq:modeI} \\
\vec{\pi}_2 &= \vec{\varphi}_2 + \vec{\varphi}_3, \label{eq:modeII} \\
\vec{\pi}_3 &= \vec{\varphi}_1 + \vec{\varphi}_2 + \vec{\varphi}_3. \label{eq:modeIII}
\end{align}
The individual modes~$\vec{\varphi}_i$, $i = 1, 2, 3$, correspond to the shear pattern~I (Figs.~\ref{fig:pattern1} and~\ref{fig:mode1}) developing perpendicular to each of the cell wall directions, i.e.~at~$\theta = 90^\circ$ and~$\pm 30^\circ$, which can be expressed in an analytical form. The first mode reads~\cite[see][Eq.~(3)]{Rokos:2019a}
\begin{equation}
\vec{\varphi}_1(\vec{X}) = \frac{1}{C_1} \left[ \sin\left(\frac{2\pi X_2}{\sqrt{3}\ell}\right)\vec{e}_1 + \frac{1}{\sqrt{3}}\sin\left(\frac{2 \pi X_1}{\ell}\right)\vec{e}_2 \right],
\label{eq:mode_1}
\end{equation}
were~$C_1$ is a normalization constant ensuring that~$\frac{1}{|Q|}\int_Q\|\vec{\varphi}_1(\vec{X})\|\,\mathrm{d}\vec{X} = 1$ for the periodic cell~$Q$ of Fig.~\ref{fig:hexab}, whereas modes~II and~III are obtained by rotating~$\vec{\varphi}_1$ by~$\mp60^\circ$, see Figs.~\ref{fig:mode2} and~\ref{fig:mode3}.

The example analysed in this section represents an infinite specimen, made of a hexagonal cellular structure with a hole diameter~$d = 1.28$~mm and a centre-to-centre spacing~$\ell = 1.386$~mm, subjected to biaxial compression. For the micromorphic computational homogenization it is modelled with a~$10 \times 10$~mm$^2$ periodic square domain discretized with eight identical quadratic triangular elements of size~$h_\mathrm{m} = 3.6\ell$ with a three-point Gauss integration rule, see Fig.~\ref{fig:hexac}. Again, the same macroscopic discretization is used for all three micromorphic fields~$v_i$, $i = 1, 2, 3$, as well as for the mean solution~$\vec{v}_0$. The RVE, shown in Fig.~\ref{fig:hexab}, discretized with isoparametric quadratic triangular elements of average size~$h_\mathrm{m} = \ell/10$ using a three-point Gauss integration rule, is assigned to each macroscopic integration point. The three orthogonal modes~$\vec{\varphi}_i$ from Figs.~\ref{fig:mode1}--\ref{fig:mode3} are considered in the ansatz in Eqs.~\eqref{eq:u} and~\eqref{eq:approxu}, i.e.~$n = 3$. A similar preliminary analysis of this case has been reported in~\cite{Rokos:2019a}, which was limited by the capabilities of the employed (quasi-Newton) solver. Here a more detailed study is presented because a full Newton solver and a bifurcation analysis are used instead.

\begin{figure}[htbp]
	\centering
	\begin{tabular}{@{}c@{}r@{\hskip 0.9em}l@{}}
	\subfloat[deformed RVE]{\includegraphics[scale=1]{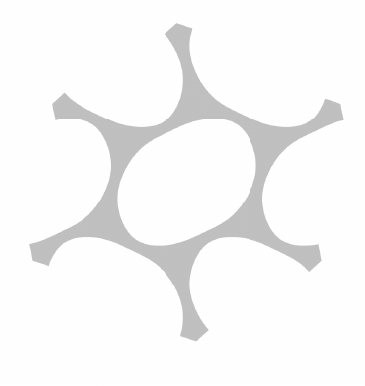}\label{micromorphic:I-IIIa}} &
	\subfloat[$\vec{e}_1$ component of~$\vec{v}_0$~\mbox{[mm]}]{\includegraphics[scale=1]{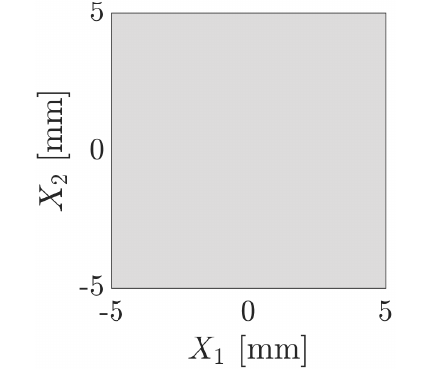}\label{micromorphic:I-IIIb}}&
	\subfloat[$\vec{e}_2$ component of~$\vec{v}_0$~\mbox{[mm]}]{\includegraphics[scale=1]{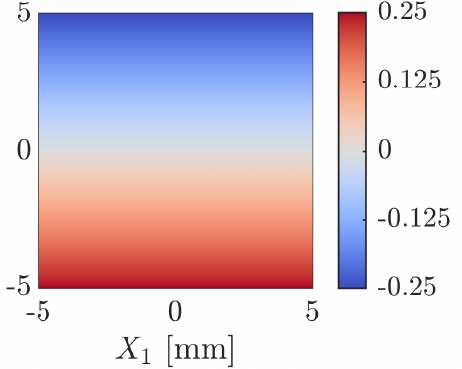}\label{micromorphic:I-IIIc}}\\ 
	\subfloat[$\widehat{v}_1$~\mbox{[-]}]{\includegraphics[scale=1]{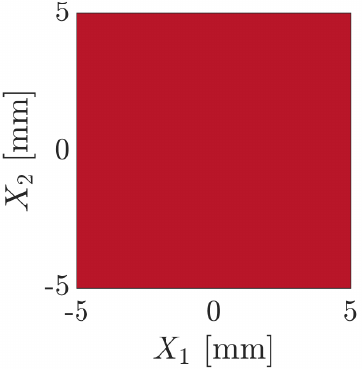}\label{micromorphic:I-IIId}}&
	\subfloat[$\widehat{v}_2$~\mbox{[-]}]{\includegraphics[scale=1]{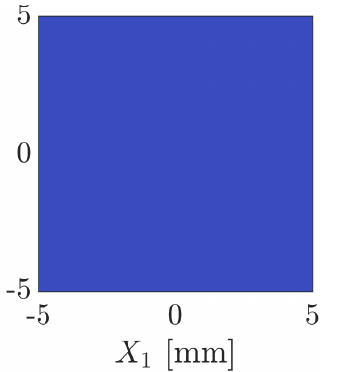}\label{micromorphic:I-IIIe}}&
	\subfloat[$\widehat{v}_3$~\mbox{[-]}]{\includegraphics[scale=1]{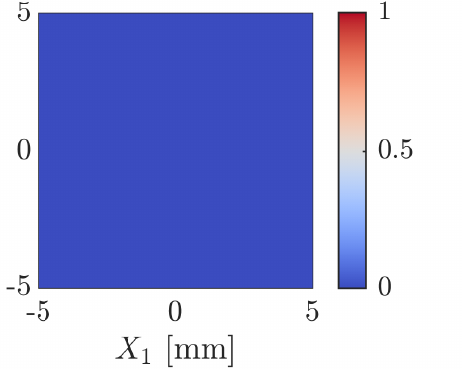}\label{micromorphic:I-IIIf}}
	\end{tabular}
	\caption{(a)~A single deformed RVE and the resulting macroscopic fields of a homogenized infinite specimen subjected to a uniaxial compressive strain of~${\varepsilon}_{22} = 0.05$ ($\gamma = \infty$). The two components of the mean displacement field~$\vec{v}_0$ are shown in~(b) and~(c), whereas the fields~$\widehat{v}_i$ that indicate the relative activation of individual modes~$\vec{\varphi}_i$, $i = 1,2,3$, are shown in~(d)--(f), where the normalization constant is~$\max_k \| v_k(t,\vec{X}) \|_\infty = 3.7$. Linear and constant effective fields with the correct pattern~I are observed.}
	\label{fig:G1}
\end{figure}

Because an infinite specimen is considered, the DNS solution directly corresponds to the behaviour of a single periodic cell (i.e.~RVE) subjected to a compressive load of biaxiality ratio~$\gamma$. Since the deformation state is periodic, the ensemble average reduces to volume averaging, easily obtained from a single microstructural translation. The question arises, however, whether the micromorphic computational homogenization is capable of reproducing such a behaviour, i.e.~yielding an affine mean field~$\vec{v}_0$ and constant micromorphic fields~$\widehat{v}_i$, while providing patterns of Eqs.~\eqref{eq:modeI}--\eqref{eq:modeIII} as the outcome of the analysis. Fig.~\ref{fig:G1} collects the results for a uniaxial compressive strain~$\varepsilon_{22} = -0.05$ ($\gamma = \infty$). As expected, the mean displacement in the~$\vec{e}_1$ direction is zero, whereas in the~$\vec{e}_2$ direction it is linear and corresponds to the applied nominal strain. The normalized micromorphic fields~$\widehat{v}_i = v_i / \max_{k=1,2,3} \| v_k(t,\vec{X}) \|_\infty$ are also spatially constant with~$\widehat{v}_1 = 1$ and~$\widehat{v}_2 = \widehat{v}_3 = 0$, resulting in an activation of mode~I and, consequently, pattern~I (recall Eq.~\eqref{eq:modeI}). Moreover, the deformed RVE shape in Fig.~\ref{micromorphic:I-IIIa} matches the DNS solution in Fig.~\ref{fig:pattern1}, corroborating further the validity of the micromorphic results.

\begin{figure}[htbp]
	\centering
	\includegraphics[scale=1]{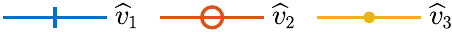}\vspace{-0.5em}\\
	\subfloat[$\gamma = \infty$]{\includegraphics[scale=1]{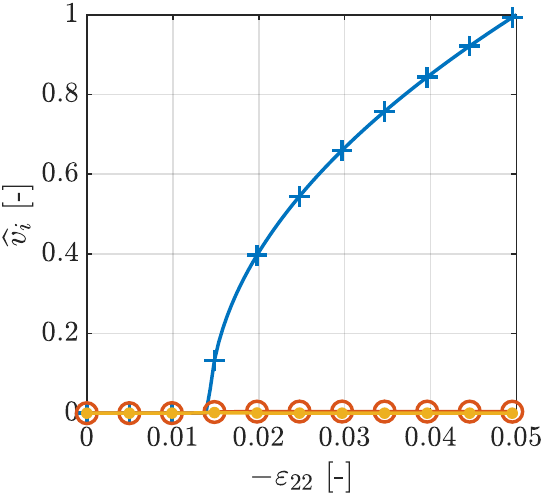}\label{fig:G2v1}}
	\hspace{0.25em}
	\subfloat[$\gamma = \frac{3}{10}$]{\includegraphics[scale=1]{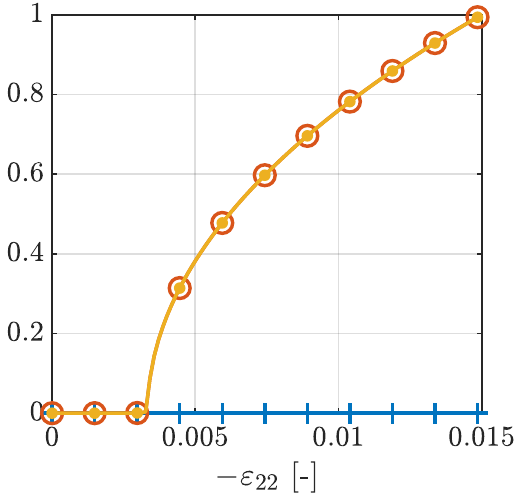}\label{fig:G2v2}}
	\hspace{0.0em}	
	\subfloat[$\gamma = 1$]{\includegraphics[scale=1]{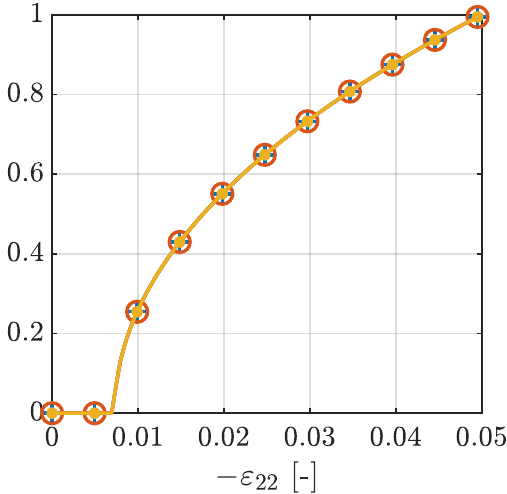}\label{fig:G2v3}}
	\caption{Magnitudes of all three spatially constant modes~$\widehat{v}_i$, $i = 1,2,3$, as a function of the overall applied vertical strain~$\varepsilon_{22}$. Three biaxiality ratios are considered: (a)~$\gamma = \infty$ (pattern~I, $-\varepsilon_{22} \in [0,0.05]$, $\varepsilon_{11} = 0$, $\max_k \| v_k(t,\vec{X}) \|_\infty = 3.7$), (b)~$\gamma = \frac{3}{10}$ (pattern~II, $-\varepsilon_{22} \in [0,0.015]$, $\varepsilon_{11} = \frac{10}{3}\varepsilon_{22}$, $\max_k \| v_k(t,\vec{X}) \|_\infty = 3.0$), and~(c) $\gamma = 1$ (pattern~III, $-\varepsilon_{11} = -\varepsilon_{22} \in [0,0.05]$, $\max_k \| v_k(t,\vec{X}) \|_\infty = 3.2$), see also Eqs.~\eqref{eq:modeI}--\eqref{eq:modeIII}.}
	\label{fig:time}
\end{figure}

The evolution of the magnitudes corresponding to the individual micromorphic fields as a function of~$\varepsilon_{22}$ is shown for the overall applied deformation gradient
\begin{equation}
\overline{\bs{F}}(t) = (1+\varepsilon_{11})\vec{e}_1\vec{e}_1+(1+\varepsilon_{22})\vec{e}_2\vec{e}_2
\label{eq:loading}
\end{equation}
and three values for~$\gamma$ in Fig.~\ref{fig:time}. Prior to bifurcation, all micromorphic fields remain zero. Upon reaching the critical strain, activation of the micromorphic fields starts exactly at the bifurcation point where a negative or sufficiently small lowest eigenvalue is observed and the system is perturbed towards the corresponding eigenvector. The correct patterns are triggered, i.e.~$\widehat{v}_1 \neq 0$ while~$\widehat{v}_2 = \widehat{v}_3 = 0$ for pattern~I ($\gamma = \infty$), $\widehat{v}_2 =  \widehat{v}_3 \neq 0$ while~$\widehat{v}_1 = 0$ for pattern~II ($\gamma = \frac{3}{10}$), and~$\widehat{v}_1 = \widehat{v}_2 = \widehat{v}_3 \neq 0$ for pattern~III ($\gamma = 1$), recall Eqs.~\eqref{eq:modeI}--\eqref{eq:modeIII}. To verify that the observed patterns in all three cases correspond to the correct solutions (i.e.~the one related to the lowest strain energy), the existence of multiple local minima is explored. To this end, all micromorphic fields are initialized as constant fields, with magnitudes spanning the entire cube~$[\widehat{v}_1,\widehat{v}_2,\widehat{v}_3] \in [0,1] \times [0,1] \times [0,1]$ for a fixed applied overall strain which corresponds to a buckled state, while assuming the exact mean fields~$\vec{v}_0$. It is found that although other patterns may yield stable local minima, the global minima always correspond to the correct combinations of modes. Note that similarly to the DNS, the multiplicities of the bifurcation points associated with the second and third pattern occur also for the micromorphic formulation. In that case, the associated buckling modes have a zero mean~$\vec{v}_0 = \vec{0}$ and spatially constant micromorphic fields, spanning the same vector space as in the DNS case. The only reliable procedure to identify the proper solution is then to explore each equilibrium path separately, opting for the one requiring the least amount of elastic strain energy. Although it may seem at this point that a bifurcation analysis is not of much benefit for a hexagonal stacking of holes, it reduces the number of possible combinations that would otherwise have to be considered as initial guesses for a quasi-Newton solver. In the case of pattern~I, the benefit is clearly substantial. For pattern~II the dimensionality reduces from three to two, whereas for pattern~III the entire space of dimensionality three should be considered. From numerical evidence, however, mode combinations approximately matching the three patterns are typically observed as eigenmodes corresponding to the three lowest eigenvalues obtained during simulations, thus reducing all possible options (spanning a vector space of dimensionality three) to only three options.
\begin{figure}[htbp]
	\centering
	\subfloat[phase diagram]{\includegraphics[scale=1]{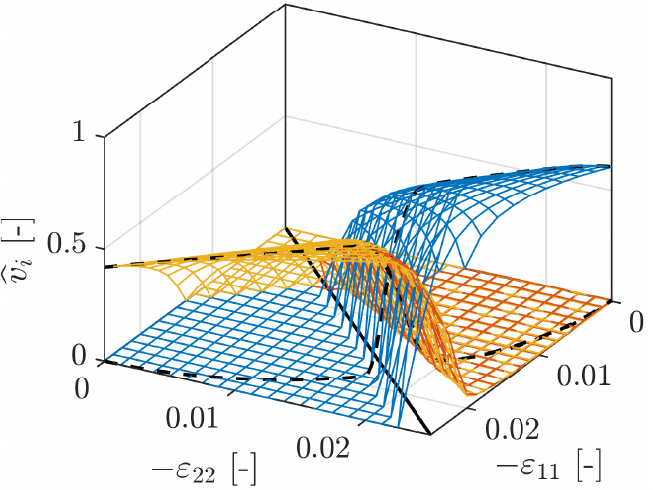}\label{fig:phaseDiagrama}}
	\hspace{0.5em}
	\subfloat[magnitude contour plot]{\includegraphics[width=5.8cm]{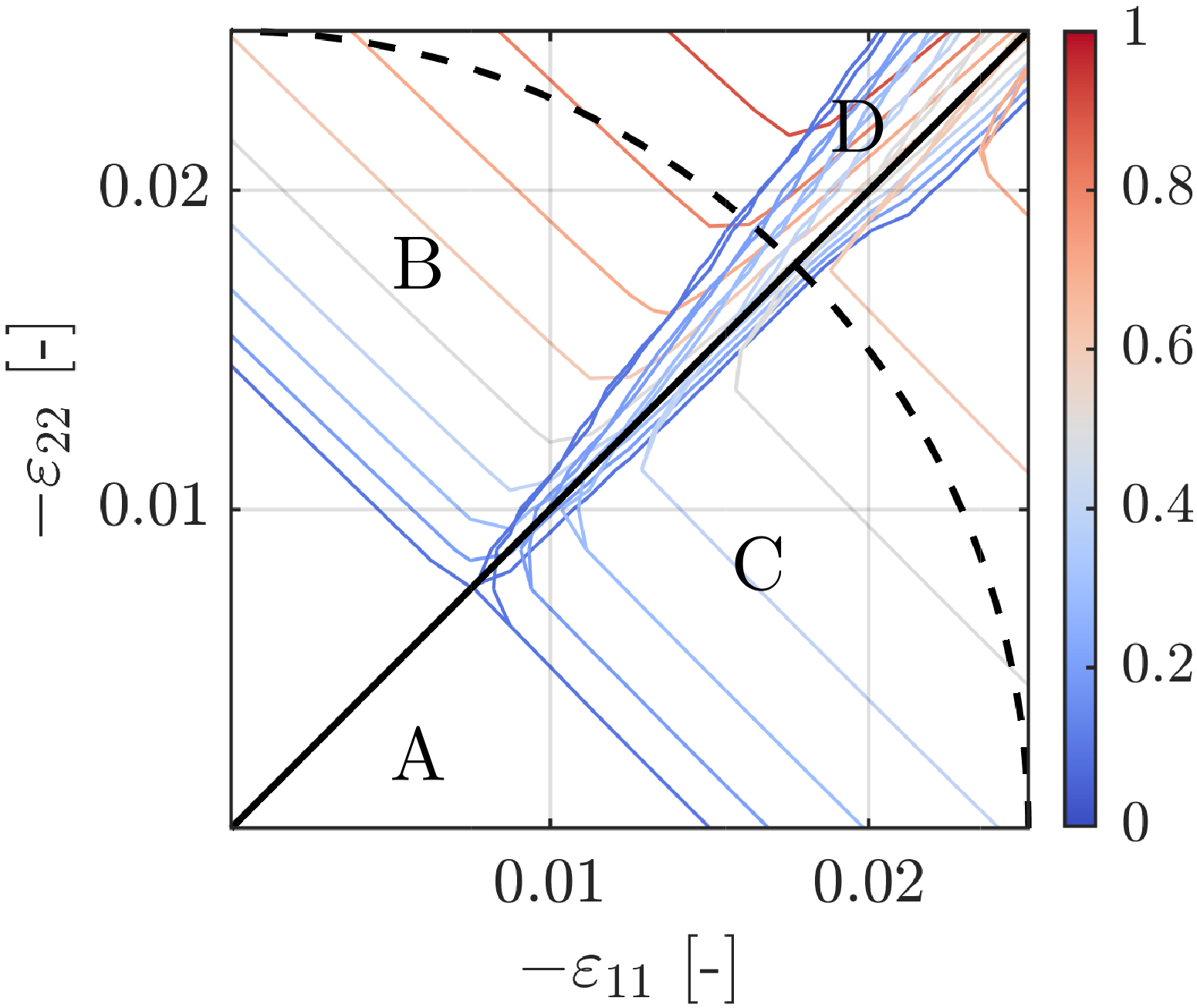}\label{fig:phaseDiagramb}} \\\vspace{1em}
	\includegraphics[scale=1]{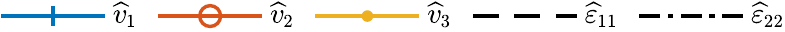}\\\vspace{-0.7em}
	\subfloat[circumferential section]{\includegraphics[scale=1]{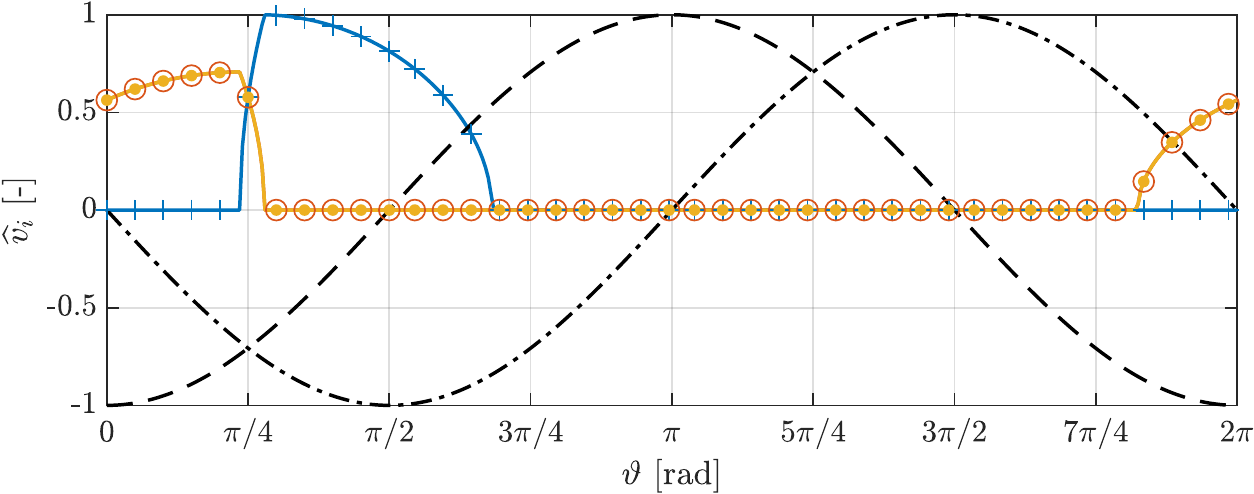}\label{fig:phaseDiagramc}}
	\caption{A phase diagram of individual normalized mode magnitudes~$\widehat{v}_i$ plotted as a function of the overall applied strains~$\varepsilon_{11}$ and~$\varepsilon_{22}$. The entire surface plot is shown in~(a) with the colour coding as used in~(c), the corresponding contour plot is shown in~(b) (normalization constant~$\max_k \| v_k(t,\vec{X}) \|_\infty = 3.3$), whereas a circumferential section for~$\widehat{\varepsilon}_{ii} = \varepsilon_{ii}/0.025$ in~(c) (normalization constant~$\max_k \| v_k(t,\vec{X}) \|_\infty = 2.8$). The circumferential section is taken along the black dashed curves in~(a) and~(b) in a clockwise direction.}
	\label{fig:phaseDiagram}
\end{figure}

The entire quadrant~$[\varepsilon_{11}, \varepsilon_{22}] \in [0,-0.025] \times [0,-0.025]$ is further explored with the micromorphic computational homogenization to provide a phase diagram of the hexagonal microstructure. The obtained result is plotted in Fig.~\ref{fig:phaseDiagrama}, where the normalized micromorphic fields~$\widehat{v}_i$ are shown. Four regions A--C are distinguished, as depicted in the corresponding contour plot in Fig.~\ref{fig:phaseDiagramb}. In the first region, A, no pattern is triggered, because the critical bifurcation strain has not been exceeded yet. Pattern~I occurs in region~B, whereas the second pattern is triggered in region~C. In the fourth region, D, a mixture of both patterns is observed, with the special configuration for~$\gamma = 1$ corresponding to pattern~III (denoted by the black solid line). A similar behaviour has been observed for hexagonal honeycomb structures in the work of~\cite{Okumura2002}; see Fig.~10 therein. Fig.~\ref{fig:phaseDiagramc} plots a circumferential section through the phase diagram of~\ref{fig:phaseDiagrama} taken along the dashed black curve highlighted in Fig.~\ref{fig:phaseDiagramb}, including its extension to the other three strain quadrants. The magnitudes of the individual normalized micromorphic fields~$\widehat{v}_i$ are plotted as a function of angle~$\vartheta \in [0, 2\pi]$, spanning the entire circle. The angle starts from the~$(\varepsilon_{22} = -1,\varepsilon_{11} = 0)$ direction and sweeps clockwise. Again, Fig.~\ref{fig:phaseDiagramc} confirms that equal magnitudes of all modes occur for~$\vartheta = \pi/4$. Furthermore, it is clearly visible which strain combinations yield which microstructural pattern. For instance, close to~$\vartheta = 5\pi/8$ and~$15\pi/8$ we notice that even though one of the applied strains~$\varepsilon_{ii}$ is positive, a pattern transformation occurs due to a large negative magnitude of the other compressive strain.
%
%
\section{Summary and Conclusions}
\label{ch:conclusions}
This contribution has extended a recently developed micromorphic computational homogenization framework for mechanical metamaterials with a full Newton solver. The micromorphic framework decomposes the kinematic field by exploiting prior knowledge on the typical patterning modes, allowing to accurately capture non-local effects present in the microstructure. The derivation and implementation of a full Newton solver for this framework has been provided, including analytical expressions for the first and second variations of the total effective potential energy. Significant gains have been obtained compared to the existing quasi-Newton implementation, in particular with respect to the bifurcation analysis, which is essential for applications of elastomeric mechanical metamaterials relying on local and global buckling. Two examples have been tested to demonstrate the capabilities of the presented numerical scheme. In the first example, a metamaterial column consisting of a microstructure with a square stacking of holes has been analysed for various slenderness ratios, for which a competition between the local and global buckling exists. The second example elaborated a uniformly loaded infinite specimen with a hexagonal stacking of holes, which may buckle into different patterns depending on the loading direction. 

The main conclusions of this paper can be summarized as follows:
\begin{enumerate}
	\item The developed full Newton solver for the micromorphic computational homogenization framework is robust and efficient.
	
	\item The micromorphic approach captures the behaviour of the reference Direct Numerical Simulation~(DNS) accurately in terms of both local and global buckling as well as the pattern magnitudes.
	
	\item The nominal stresses are reproduced by the micromorphic framework with a good accuracy, although the post-bifurcation results are in general systematically overestimated compared to DNS. The maximum error in terms of the critical buckling stress corresponding to the first instability point does not exceed~$12\%$, and decreases with increasing scale ratio down to approximately~$7\%$.

	\item The buckling strain is captured with a higher accuracy compared to the nominal buckling stress. The relative error stays below~$3\%$ for local and~$6\%$ for global buckling, and decreases down to~$0.5\%$ for local and~$3\%$ for global buckling with increasing scale ratio.
	
	\item The micromorphic scheme reproduces DNS results correctly even in the case of a hexagonal stacking of holes, for which multiple patterning modes occur. It predicts the correct patterns for the loading directions considered.
\end{enumerate}

The full Newton solver presented here greatly reduces the dependency of the solution on the initial guess by perturbing the system towards the correct direction when a bifurcation point is encountered; therefore, it provides an indispensable numerical tool for modelling instability-based mechanical metamaterials.
%
%
%
%
\section*{Acknowledgements}
The research leading to these results has received funding from the European Research Council under the European Union's Seventh Framework Programme (FP7/2007-2013)/ERC grant agreement no.~[339392] (O. Roko\v{s} 09/2016--03/2019, R.H.J. Peerlings, and M.G.D Geers) and from the Czech Science Foundation (GA\v{C}R) grant agreement no.~[19-26143X] (O. Roko\v{s} 03/2019--12/2019, and M. Do\v{s}k\'{a}\v{r}). The authors would like to also acknowledge Prof.~Jan Zeman from the Czech Technical University in Prague for fruitful discussions and critical comments on the manuscript, and Dr.~Geralf H\"{u}tter from TU Bergakademie Freiberg for valuable email communication.
%
%
%
\bibliography{mybibfile}
\end{document}